\documentclass[journal, 10pt]{IEEEtran}

\usepackage{afterpage}
\usepackage{algorithmicx}
\usepackage[linesnumbered,ruled,noend]{algorithm2e}
\SetKwInOut{Input}{Input}
\SetKwInOut{Output}{Output}

\SetKwProg{Fn}{Function}{\string:}{end}

\usepackage[backend=biber, natbib=true, style=ieee, url=false, doi=false, sorting=none, maxnames=3, minnames=3]{biblatex}

\usepackage[a4paper, margin=1.82cm]{geometry}

\usepackage{lipsum}
\usepackage{amsfonts}
\usepackage{amsmath}
\usepackage{amssymb}
\usepackage{array}
\usepackage{bm}
\usepackage{booktabs}
\usepackage{boxedminipage}
\usepackage{colortbl}
\usepackage{enumerate}
\usepackage{extarrows}
\usepackage{float}
\usepackage{graphics}
\usepackage{graphicx}
\usepackage{hyperref}
\usepackage{indentfirst}
\usepackage{latexsym}
\usepackage{makecell}
\usepackage{mismath}
\usepackage{multirow}
\usepackage{picinpar}
\usepackage{placeins}
\usepackage{subfigure}
\usepackage{textcomp}
\usepackage{theorem}
\usepackage[para]{threeparttable}
\usepackage[normalem]{ulem}
\usepackage{url}
\usepackage{wrapfig}
\usepackage{xcolor}

\newtheorem{example}{Example}

\IEEEoverridecommandlockouts

\allowdisplaybreaks

\usepackage{threeparttable}

\newcommand{\tit}[1]{{\textit{#1}}}
\newcommand{\tbf}[1]{{\textbf{#1}}}
\newcommand{\ie}{\textit{i.e.}}
\newcommand{\eg}{\textit{e.g.}}

\newcommand{\cadd}[1]{{\color{black}{#1}}} 
\newcommand{\cdel}[1]{}
\newcommand{\radd}[1]{{\color{black}{#1}}} 
\newcommand{\rdel}[1]{}
\newcommand{\fvdel}[1]{}

\urldef{\abcslides}\url{https://ethz.ch/content/dam/ethz/special-interest/itet/efcl-dam/documents/Introduction%20to%20ABC.pdf}

\begin{document}

\title{Simulation-Guided Approximate Logic Synthesis Under the Maximum Error Constraint}

\author{Chang~Meng,~\IEEEmembership{Member, IEEE},
Weikang~Qian,~\IEEEmembership{Senior~Member, IEEE}, 
and Giovanni~De~Micheli,~\IEEEmembership{Life~Fellow, IEEE}
\thanks{Chang~Meng is with the Integrated Systems Laboratory, École Polytechnique Fédérale de Lausanne, Switzerland (email: chang.meng@epfl.ch).}
\thanks{Weikang~Qian is with the Global College, Shanghai Jiao Tong University, China (email: qianwk@sjtu.edu.cn).}
\thanks{Giovanni~De~Micheli is with the Integrated Systems Laboratory, École Polytechnique Fédérale de Lausanne, Switzerland (email: giovanni.demicheli@epfl.ch).}
\thanks{Corresponding author: Chang Meng.}}

\maketitle

\begin{abstract}
Approximate computing is an effective computing paradigm for improving the energy efficiency of error-tolerant applications.
Approximate logic synthesis (ALS) is an automatic process to generate approximate circuits with reduced area, delay, and power, while satisfying user-specified error constraints.
This paper focuses on ALS under the maximum error constraint.
As an essential error metric that provides a worst-case error guarantee,
the maximum error is crucial for many applications such as image processing and machine learning.
This work proposes an efficient simulation-guided ALS flow that handles this constraint.
It utilizes logic simulation to 1) prune local approximate changes (\tit{LACs}) with large errors that violate the error constraint,
and 2) accelerate the SAT-based LAC selection process.
Furthermore, to enhance scalability,
our ALS flow iteratively selects a set of promising LACs satisfying the error constraint to improve \fvdel{the }efficiency.
The experimental results show that compared with the state-of-the-art method, 
our ALS flow accelerates by $30.6\times$,
and further reduces the circuit area and delay by $18.2\%$ and $4.9\%$, respectively.
\radd{Notably, our flow scales to large EPFL benchmarks with up to 38540 nodes,
which \rdel{cannot be handled by any existing ALS method for maximum error}remain challenging for existing ALS methods tackling maximum error constraint.}

\end{abstract}


\section{Introduction}\label{sec:intr}

Approximate computing~\cite{han2013approximate} is an emerging low-power design paradigm for error-tolerant applications
such as signal processing, data mining, and machine learning~\cite{mittal2016survey}.
It carefully introduces errors to significantly reduce the hardware cost,
while the application-level quality is almost unaffected.
\emph{Approximate logic synthesis} (\emph{ALS}) is an automatic process to generate approximate circuits~\cite{scarabottolo2020approximate}.
An ALS tool takes an accurate circuit and user-specified error constraints as inputs
and outputs an approximate circuit with smaller area, delay, and power, satisfying the constraints.

To evaluate the accuracy of an approximate circuit,
two types of error metrics
are utilized, the average error and maximum error~\cite{scarabottolo2020approximate}.
Average error, such as error rate and mean error distance, measures the average deviation between the outputs of the accurate and approximate circuits, while the maximum error measures the maximum deviation between the outputs of the accurate and approximate circuits over all input patterns.
Typical maximum errors include \emph{maximum error distance} (\emph{MaxED}) and \emph{maximum Hamming distance} (\emph{MaxHD}).
Maximum error provides a worst-case guarantee of the error,
which is crucial for many applications.
For instance, in image processing, even if the average error is low, 
occasional large errors can lead to visible artifacts.
Another example is machine learning, 
where rare extreme deviations may cause mispredictions.
Using maximum error constraints ensures that every computation remains within safe bounds, 
preserving overall quality and reliability.
Considering the importance of maximum error, this paper focuses on the ALS under the maximum error constraint.

Many ALS methods for maximum error constraint have been proposed~\cite{venkataramani2012salsa, scarabottolo2018circuit, witschen2022muscat, rezaalipour2023parametrizable},
most of which simplify the circuit by applying \emph{local approximate changes} (\emph{LACs}).
A LAC is a local modification of the circuit.
For example, a \tit{constant LAC}~\cite{shin2011new} replaces a signal by a constant 0 or 1,
and a \tit{SASIMI LAC}~\cite{venkataramani2013substitute} substitutes a signal by another.
After generating candidate LACs,
an ALS flow estimates the maximum error caused by the LACs.
Based on the error estimation results,
the ALS flow then identifies LACs that can be applied to simplify the circuit while satisfying the error constraint.
There are two categories of maximum error estimation methods.
The first category estimates an upper bound of the maximum error~\cite{schlachter2017design, scarabottolo2018circuit, scarabottolo2019partition, scarabottolo2021formal}.
However, these methods only support the simple constant LAC~\cite{shin2011new},
while complex LACs such as the SASIMI LAC~\cite{venkataramani2013substitute} that can achieve better approximate circuits are not supported.
To handle complex LACs,
the second category checks whether the maximum error is within a user-specified bound or not~\cite{venkatesan2011macaco, chandrasekharan2016approximation, vcevska2017approximating}.
For each LAC,
these methods convert the maximum error checking problem for the LAC into a SAT problem,
and the SAT solving result determines whether the maximum error caused by the LAC is within the bound or not.
However, a new challenge is the massive number of complex LACs in a circuit.
The solving of their corresponding numerous SAT problems is time-consuming and limits the scalability of the ALS flow.
For example, for a circuit with $N$ nodes,
there are $O(N^2)$ SASIMI LACs~\cite{venkataramani2013substitute} that replace a signal by another,
and solving the corresponding $O(N^2)$ SAT problems is impractical for large circuits.

To address the above challenges,
we propose an efficient logic simulation-guided ALS flow under the maximum error constraint,
which can handle complex LACs and scale to large circuits.
Logic simulation has shown its effectiveness in accelerating the SAT solving process in many traditional logic synthesis works such as~\cite{lee2021simulation},
and this motivates us to leverage it to accelerate the maximum error checking process in ALS.
Our main contributions are as follows:
\begin{enumerate}
    \item We propose to utilize logic simulation to prune large-error LACs violating the maximum error constraint,
    which significantly reduces the number of LACs to be considered,
    and hence accelerates the ALS flow.
    \item For the remaining LACs after pruning, 
    we propose to use simulation-guided SAT solving to further accelerate the LAC selection process.
    \item Based on the simulation-based LAC pruning and fast LAC selection,
    we design an ALS flow for maximum error constraint that iteratively applies a set of promising LACs for efficient circuit simplification.
\end{enumerate}

The experimental results show that compared with the state-of-the-art method,
our ALS flow accelerates by $30.6\times$,
and further reduces the circuit area and delay by $18.2\%$ and $4.9\%$, respectively.
Our method scales to large EPFL benchmarks with up to 38540 nodes,
which \rdel{cannot be handled by any existing ALS method for maximum error constraint}%
remain challenging for existing ALS methods tackling the maximum error constraint.
Our work is open-source and available at \url{https://github.com/changmg/SimALS-MaxError}.


The remainder of this paper is organized as follows.
Section~\ref{sec:rel} reviews the related works.
Section~\ref{sec:backgr} introduces the background.
Section~\ref{sec:method} elaborates the proposed simulation-guided ALS flow under the maximum error constraint.
The experimental results are presented in Section~\ref{sec:res}, 
followed by conclusions in Section~\ref{sec:concl}.
\section{Related Works}\label{sec:rel}

This section reviews the related works on ALS methods and maximum error estimation methods for ALS.

\radd{
\subsection{Approximate Logic Synthesis (ALS) Methods}

As mentioned before,
the average error and the maximum error are two typical error metrics used in ALS.
Some ALS methods can handle both average and maximum error constraints,
such as~\cite{shin2011new} and~\cite{chandrasekharan2016approximation},
some focus on the average error constraint,
such as~\cite{venkataramani2013substitute,wu2016efficient,hashemi2018blasys,zhou2018dals,meng2023hedals,meng2020alsrac,wang2023dasals},
and \fvdel{the }others deal with the maximum error constraint,
such as~\cite{venkataramani2012salsa,scarabottolo2018circuit,scarabottolo2018circuit,witschen2022muscat,rezaalipour2023parametrizable}.
Our work differs from the above works on ALS for maximum error constraint in two aspects:
1) supporting complex LACs for better circuit quality,
and 2) using effective simulation-guided techniques to enhance scalability.

\subsection{Error Estimation Methods for ALS}

Error estimation of LACs is a critical step in ALS.
The average error estimation is usually based on Monte Carlo simulation~\cite{su2022vecbee,meng2022seals,wang2023accals} or analytical methods~\cite{echavarria2020probabilistic}.
However, these methods cannot be directly applied for maximum error estimation.

\radd{To estimate the maximum error caused by LACs, two categories of methods are proposed.}
The first category estimates a maximum error upper bound,
\radd{such as~\cite{schlachter2017design,scarabottolo2021formal}.}
However, both~\cite{schlachter2017design} and~\cite{scarabottolo2021formal} only consider a simple LAC that replaces a signal by a constant 0 or 1~\cite{shin2011new}.
Complex LACs such as the SASIMI LAC~\cite{venkataramani2013substitute} that can further simplify the circuit are not supported.

The second category directly checks whether the maximum error caused by each LAC is within the error bound,
\radd{such as~\cite{venkatesan2011macaco,chandrasekharan2016approximation,vcevska2017approximating,witschen2022muscat,meng2023mecals}.}
\radd{For example,}
the MUSCAT ALS flow~\cite{witschen2022muscat} encodes the maximum error checking of all candidate LACs into a single \tit{minimal unsatisfiable subset} (\tit{MUS}) problem.
The solution of the MUS problem corresponds to an optimized approximate circuit satisfying the maximum error constraint.
However, solving the MUS problem is very time-consuming, 
and MUSCAT does not support complex LACs like the SASIMI LAC~\cite{venkataramani2013substitute}.
Although MUSCAT sets a time limit for the MUS solving to enhance scalability,
the time limit may lead to suboptimal approximate circuits.
Another approach, the MECALS ALS flow~\cite{meng2023mecals}, 
converts the maximum error checking for all candidate LACs into a SAT sweeping problem.
Unfortunately, SAT sweeping also has a scalability issue,
making MECALS unable to handle large circuits in practice.
\cdel{In our ALS flow, we use}Our method, instead, uses logic simulation to accelerate the check of maximum errors caused by numerous LACs,
which belongs to the second category of maximum error estimation methods.
Unlike the existing works,
our method can handle complex LACs and has better scalability.
}
\section{Background}\label{sec:backgr}

\cdel{This section introduces the basic concepts of logic circuit terminologies, maximum error metrics, error miter, and LACs.}%

\subsection{Logic Circuit Terminologies}

Our study focuses on multi-level combinational logic circuits,
which can be modeled as directed acyclic graphs.
For simplicity,
we use the term \emph{circuit} to refer to a multi-level combinational logic circuit.
In a circuit,
the inputs and outputs of a node are called its \emph{fanins} and \emph{fanouts}, respectively.
A \emph{primary input} (\emph{PI}) is a node without any fanin.
A \emph{functional node} is one performing a logic operation.
A \emph{primary output} (\emph{PO}) is a dummy node driven by either a functional node or a PI;
it has a single fanin and no fanouts.
A \emph{path} is a sequence of connected nodes in the circuit.
If there exists a path from node $u$ to $v$,
then $v$ is a \emph{transitive fanout} (\emph{TFO}) of $u$.

\subsection{Maximum Error Metrics}\label{subsec:error-metrics}

Consider two multiple-output Boolean functions $\bm{y}: \mathbb{B}^I\rightarrow \mathbb{B}^O$ for an accurate circuit $G$
and $\bm{\hat y}: \mathbb{B}^I\rightarrow \mathbb{B}^O$ for its approximate counterpart $\hat G$,
where $I$ and $O$ are the numbers of PIs and POs, respectively.
The maximum error of circuit $\hat G$ quantifies the maximum deviation between $\bm y$ and $\bm{\hat y}$ over all PI patterns $\bm x$ as follows:
\begin{equation}\label{eq:max_err}
    \tit{MaxError}(G) = \max_{\bm x\in \mathbb B^I} D\left(\bm y(\bm x), \bm{\hat y}(\bm x)\right),
\end{equation}
where $\bm y(\bm x)$ and $\bm{\hat y}(\bm x)$ are binary vectors of length $O$, 
denoting the PO values of $G$ and $\hat G$ under the PI pattern $\bm x$, respectively.
The function $D$ is called a deviation function,
measuring the deviation between $\bm y$ and $\bm{\hat y}$.

Typical maximum errors include \tit{maximum error distance} (\tit{MaxED}) and \tit{maximum Hamming distance} (\tit{MaxHD}).
MaxED measures the maximum absolute difference between the numerical values encoded by the POs of the accurate and approximate circuits.
Its deviation function is
\begin{equation}\label{eq:dev_maxed}
D_{\text{MaxED}}(\bm y, \bm{\hat y}) = \left| \tit{int}(\bm y) - \tit{int}(\bm{\hat y}) \right|,
\end{equation}%
where $\tit{int}(\bm v)$ returns the integer encoded by the binary vector $\bm v$.
For example, if $\bm y$ encodes an $O$-bit unsigned integer,
then $\tit{int}(\bm y) = \sum_{k=1}^{O} 2^{k-1} y_k$,
where $y_k$ denotes the $k$-th bit of the binary vector $\bm y$.
By measuring the numerical deviation,
MaxED is a suitable metric for arithmetic circuits, such as adders and multipliers.

MaxHD measures the maximum number of bit-flips between $\bm y$ and $\bm{\hat y}$.
Its deviation function is 
\begin{equation}\label{eq:dev_maxhd}
D_{\tit{MaxHD}}(\bm y, \bm{\hat y}) = \sum_{k=1}^{O} \left| y_k - \hat y_k \right|.
\end{equation}%
By limiting the number of bit-flips,
MaxHD is a suitable metric for digital communication and error correction circuits.

\subsection{Error Miter for Maximum Error Checking}\label{subsec:approx_miter}

\begin{figure}[!htbp]
    \centering
    \includegraphics[width=1.0\columnwidth]{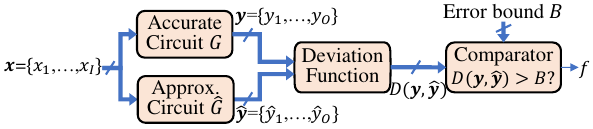}
    \caption{An error miter that checks whether the maximum error of the approximate circuit $\hat G$ exceeds the error bound $B$ or not.}\label{fig:error_miter}
    \vspace{-1.0em}
\end{figure}

An error miter is an auxiliary circuit to check 
whether the maximum error of an approximate circuit exceeds a given bound or not~\cite{venkatesan2011macaco,wu2019alfans}. 
As shown in Fig.~\ref{fig:error_miter},
it consists of an accurate circuit $G$,
an approximate circuit $\hat G$, a deviation function unit, and a comparator.
The accurate and approximate circuits take the same PIs $\bm x$ as inputs,
and their corresponding outputs are $\bm y$ and $\bm{\hat y}$, respectively.
The deviation unit computes the deviation function $D(\bm y, \bm{\hat y})$,
such as Eq.~\eqref{eq:dev_maxed} or Eq.~\eqref{eq:dev_maxhd}.
The comparator checks whether $D(\bm y, \bm{\hat y})$ is larger than the bound $B$.
If $D(\bm y, \bm{\hat y}) > B$, the output of the comparator, $f$, is $1$; otherwise, $f$ is $0$.

To check the maximum error of circuit $\hat G$,
the error miter is converted into a SAT problem.
If the solver returns SAT, 
then there exists a PI pattern $\bm x$ causing $f = 1$.
In this case, we have $D(\bm y, \bm{\hat y}) > B$,
and hence $\tit{MaxError}(\hat G) > B$.
Otherwise, if the solution is UNSAT, 
then $f$ is always $0$.
This means that $D(\bm y, \bm{\hat y}) \le B$ over all PI patterns $\bm x$,
implying $\tit{MaxError}(\hat G) \le B$.

\section{Methodology}\label{sec:method}

This section introduces our simulation-guided ALS flow under the maximum error constraint.
We first overview the flow in Section~\ref{subsec:overview},
followed by the details in Sections~\ref{subsec:prune} and~\ref{subsec:sat}.

\subsection{Overview}\label{subsec:overview}

Our ALS flow aims to solve this problem:
Given an accurate circuit $G$ in any graph representation (\eg{}, AIG, gate netlist, etc.)
and a maximum error bound $B$,
find a min-area approximate gate netlist $G_\tit{final}$,
while ensuring $\tit{MaxError}(G_\tit{final}) \leq B$.

\begin{figure}[!htbp]
    \centering
    \includegraphics[width=0.70\columnwidth]{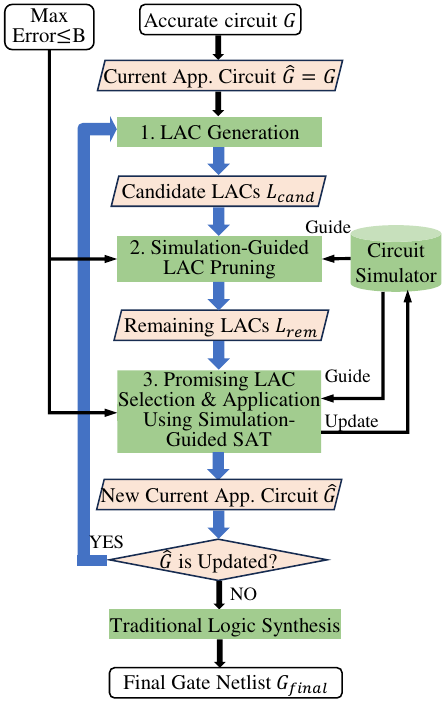}
    \vspace{-1.0em}
    \caption{Simulation-guided ALS flow under the maximum error constraint.}
    \label{fig:ALS_flow}
    \vspace{-1.0em}
\end{figure}

As shown in Fig.~\ref{fig:ALS_flow}, our
flow starts by initializing a \tit{current approximate circuit} $\hat G$ as a copy of the accurate circuit $G$.
Then, circuit $\hat G$ is iteratively simplified in a main loop,
indicated by the blue arrows in Fig.~\ref{fig:ALS_flow}.
%
Each iteration consists of three key steps.
Step 1 generates a set of candidate LACs $L_{\tit{cand}}$.
Here, the generated LACs can be any \emph{single-output LACs},
\textit{i.e.}, LACs whose affected local circuits have only one output,
such as constant LACs and SASIMI LACs.
Since $L_\tit{cand}$ usually contains numerous LACs,
step 2 prunes the invalid LACs violating the error constraint according to logic simulation results.
The set of remaining LACs after pruning is denoted as $L_\tit{rem}$.
Then, step 3
selects a set of promising LACs from $L_\tit{rem}$ and applies them to simplify the current approximate circuit $\hat G$,
where a promising LAC refers to a LAC whose application significantly reduces circuit area while satisfying the error constraint.
If the approximate circuit $\hat{G}$ is successfully simplified compared to the previous iteration, the main loop continues for the next iteration.
Otherwise, no more valid LACs exist,
and the main loop terminates.
Then, traditional logic synthesis is performed to further simplify the circuit $\hat G$ without introducing additional errors, 
producing the final approximate gate netlist $G_\tit{final}$.

Note that logic simulation serves as a guider in our ALS flow.
As shown in the middle right part of Fig.~\ref{fig:ALS_flow},
the circuit simulator not only guides the pruning of invalid LACs in $L_\tit{cand}$,
but also accelerates the LAC selection process by guiding the SAT solving.
Furthermore, the simulation patterns in the circuit simulator are updated on the fly by the LAC selection information from step 3.
This technique further accelerates the ALS flow by reducing the number of SAT problems to be solved.

The following subsections detail the key steps in our ALS flow.
Specifically, Section~\ref{subsec:prune} introduces step 2, the simulation-guided LAC pruning,
and Section~\ref{subsec:sat} describes step 3, the selection and application of the promising LACs using simulation-guided SAT solving.

\subsection{Simulation-Guided LAC Pruning}\label{subsec:prune}
The LAC pruning step filters out some invalid candidate LACs in $L_\tit{cand}$ generated in step 1 of our flow
and returns a set of remaining LACs, denoted as $L_\tit{rem}$.
There are usually many LACs in $L_\tit{cand}$.
For example, a circuit with $N$ nodes has $O(N^2)$ SASIMI LACs~\cite{venkataramani2013substitute} that replace a node by another.
If we use the error miter-based method (see Section~\ref{subsec:approx_miter}) to check the maximum error of each candidate LAC in $L_\tit{cand}$,
then $O(N^2)$ SAT problems need to be solved,
which is impractical for large circuits.
\rdel{Given that simulation is typically much faster than SAT solving}\radd{Given that simulating a small subset of all possible input patterns (\eg{}, $\le 2^{13}$ patterns in our experiments) is typically much faster than SAT solving},
we propose to use simulation to quickly prune the invalid LACs in $L_\tit{cand}$ violating the maximum error constraint.
After the pruning,
the number of LACs to be checked by SAT solving is significantly reduced,
thus improving the efficiency of the ALS flow.

In the following parts,
we first introduce a theoretical foundation of the simulation-guided LAC pruning in Section~\ref{subsubsec:prune_theory},
\ie{}, simulation can obtain a lower bound of the maximum error caused by a candidate LAC.
Next, we describe how to efficiently compute the lower bound on the maximum error in Section~\ref{subsubsec:prune_compute}.

\subsubsection{Theoretical Foundation}\label{subsubsec:prune_theory}
\radd{Given a candidate LAC $l_\tit{cand}$, 
logic simulation over a subset of PI patterns yields the maximum observable error under these patterns,
denoted as $\tit{MaxError}_\tit{LB}(l_\tit{cand})$.
Since the simulation does not cover all PI patterns,
this value is a lower bound on the real maximum error caused by $l_\tit{cand}$, \ie{},
\begin{equation}\label{eq:maxerr_lb}
\tit{MaxError}_\tit{LB}(l_\tit{cand}) \leq \tit{MaxError}_\tit{real}(l_\tit{cand}).
\end{equation}

This property forms the basis for the simulation-guided LAC pruning.
If simulation finds that $\tit{MaxError}_\tit{LB}(l_\tit{cand})\!>\!B$,
then $l_\tit{cand}$ is guaranteed to violate the maximum error constraint and is discarded.}
In our implementation,
we first simulate with a small number of $M_\tit{small}$ random PI patterns to roughly prune the invalid LACs,
followed by a larger number of $M$ PI patterns to obtain a tighter lower bound of the maximum error for further \cdel{prune the invalid LACs}pruning.
The LACs not pruned by the two-round simulation are retained in $L_\tit{rem}$.

\subsubsection{Efficient Computation of the Maximum Error Lower Bounds}\label{subsubsec:prune_compute}

The theoretical foundation of the simulation-guided LAC pruning in Eq.~\eqref{eq:maxerr_lb} requires computing $\tit{MaxError}_\tit{LB}(l_\tit{cand})$ for each candidate LAC $l_\tit{cand}$.
To compute $\tit{MaxError}_\tit{LB}(l_\tit{cand})$ for each candidate LAC $l_\tit{cand}$,
a naive way is to apply $l_\tit{cand}$ to the current approximate circuit $\hat G$ and obtain a new approximate circuit $\hat G_\tit{cand}$.
After simulating the accurate circuit $G$ and the approximate circuit $\hat G_\tit{cand}$ under \fvdel{the PI patterns in $\mathbb S$}some sampled PI patterns,
$\tit{MaxError}_\tit{LB}(l_\tit{cand})$ can be obtained.
This method is straightforward but slow,
requiring $O(|L_\tit{cand}|)$ simulation runs,
where $|L_\tit{cand}|$ is the number of candidate LACs.

Instead of using the naive method,
we accelerate the computation of all $\tit{MaxError}_\tit{LB}(l_\tit{cand})$'s based on the change propagation matrix (CPM) proposed in~\cite{su2022vecbee}.
The CPM $P$ for the current approximate circuit $\hat G$
is a three-dimensional 0-1 matrix of size $M \times N \times O$,
where $M$ is the number of simulation patterns,
$N$ is the number of functional nodes in the circuit,
and $O$ is the number of POs in the circuit.
Each entry in the CPM is indexed as $P[i, n, \hat y_k]$,
where $1\!\le\! i\!\le\! M$ represents the $i$-th simulation pattern,
$n$ is a functional node in the circuit,
and $\hat y_k$ ($1\!\le\! k\!\le\! O$) is the $k$-th PO in $\hat G$.
The entry $P[i, n, \hat y_k]$ evaluates the impact of the change in $n$'s value on $\hat y_k$.
Specifically,
$P[i, n, \hat y_k] = 1$ indicates that a flip of $n$'s value will cause a flip of $y_k$'s value under the $i$-th pattern,
while $P[i, n, \hat y_k] = 0$ means that $\hat y_k$'s value keeps unchanged after a flip of $n$'s value under the $i$-th pattern.
To compute $P[i, n, \hat y_k]$, 
under the $i$-th pattern
we can flip $n$'s value,
update the values of all $n$'s TFOs and POs using $n$'s new value,
and then check whether $\hat y_k$'s value changes.
If it changes, then $P[i, n, \hat y_k]$ is 1; 
otherwise, $P[i, n, \hat y_k]$ is 0.
We apply the above process to each functional node $n$ in the circuit to obtain its CPM entries.
Thus, computing the CPM for all functional nodes requires $O(N)$ simulation runs,
where $N$ is the number of functional nodes.

\begin{figure}[!htbp]
    \centering
    \includegraphics[width=1.0\columnwidth]{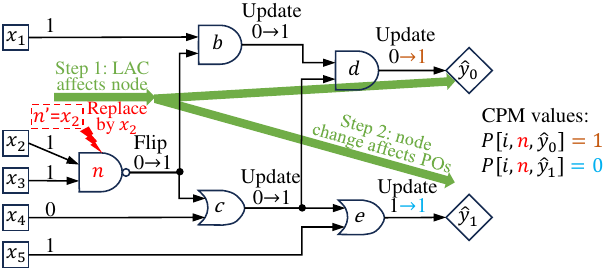}
    \vspace{-2.0em}
    \caption{An example circuit.
    The number above each wire is the signal value under the $i$-th input pattern in the simulation.
    The impact of the LAC that replaces node $n$ with node $x_2$ is considered.}
    \label{fig:cpm}
\end{figure}

\begin{example}\label{ex:cpm}
In the example circuit shown in Fig.~\ref{fig:cpm},
assume that the $i$-th PI pattern for simulation is $x_1 x_2 \ldots x_5 = 11101$.
The simulation values of the gates are shown above the wires.
To compute $P[i, n, \hat y_0]$ and $P[i, n, \hat y_1]$, we flip node $n$'s value from 0 to 1,
and update the values of $n$'s TFOs.
That is, the value of $b$ changes from 0 to 1, the value of $c$ changes from 0 to 1, 
the values of $d$ and $\hat y_0$ change from 0 to 1, 
and the values of $e$ and $\hat y_1$ keep unchanged.
Since under the $i$-th pattern, the flip of $n$'s value causes a flip of $\hat y_0$'s value, while it does not affect $\hat y_1$'s value,
we have $P[i, n, \hat y_0]= 1$ and $P[i, n, \hat y_1]= 0$.
\end{example}

CPM can be used to efficiently compute the PO values after applying each candidate LAC,
and hence the lower bound on the maximum error of the candidate LAC.
For the single-output LACs considered in this work,
each of them, denoted as $l_\tit{cand}$, can be modeled as replacing an existing node $n$ in the circuit with a new node $n'$.
After applying $l_\tit{cand}$ to the current approximate circuit $\hat G$,
we can obtain the new value of the $k$-th PO under the $i$-th pattern,
denoted as $y'_k[i]$, as follows:%
\footnote{\radd{Eq.~\eqref{eq:po_update} is an intermediate result derived from~\cite{su2022vecbee}.
However, \cite{su2022vecbee} studies the ALS problem under the average error constraint.
Here, we extend its use to the ALS problem under the maximum error constraint,
where it serves as a key building block of our simulation-guided ALS flow.}}
\begin{equation}\label{eq:po_update}
\begin{split}
    y'_k[i] & = \hat y_k[i] \oplus \tit{Impact}[i, l_\tit{cand}, \hat y_k] \\
    & = \hat y_k[i] \oplus \left((n[i] \oplus n'[i]) \land P[i, n, \hat y_k]\right), \\
\end{split}
\end{equation}%
where $\hat y_k[i]$ is the $k$-th PO's value under the $i$-th pattern before applying $l_\tit{cand}$,
$\tit{Impact}[i, l_\tit{cand},\hat y_k]$ is a binary value called the \emph{impact factor} for evaluating the impact of applying $l_\tit{cand}$ on the $k$-th PO under the $i$-th pattern, which will be explained next,
$n[i]$ and $n'[i]$ are the values of node $n$ and $n'$ under the $i$-th pattern, respectively,
and $\oplus$ and $\land$ are the XOR and AND operations, respectively.

Note that Eq.~\eqref{eq:po_update} has an impact factor $\tit{Impact}[i, l_\tit{cand}, \hat y_k]$,
measuring whether applying $l_\tit{cand}$ changes the value of $k$-th PO ($\hat y_k$) under the $i$-th pattern.
The impact factor has two components, corresponding to the two steps of the impact of $l_\tit{cand}$ on the PO, indicated by the green arrows in Fig.~\ref{fig:cpm}.
Step 1 is that the application of $l_\tit{cand}$ changes node $n$'s value under the $i$-th pattern,
which is captured by $n[i] \oplus n'[i]$.
If the value of the new node $n'$ is different from that of the original node $n$ under the $i$-th pattern,
then $n[i] \oplus n'[i]$ is 1, indicating the change of $n$'s value after applying $l_\tit{cand}$.
Otherwise, $n[i] \oplus n'[i]$ is 0, indicating that applying $l_\tit{cand}$ does not change node $n$'s value,
and hence does not affect $\hat y_k$'s value under the $i$-th pattern.
Step 2 is that the change of $n$'s value causes the change of $\hat y_k$'s value under the $i$-th pattern,
which is captured by the CPM entry $P[i, n, \hat y_k]$.
It is obvious that only if the two steps both occur,
$\hat y_k$'s value changes after applying $l_\tit{cand}$ under the $i$-th pattern.
Therefore, the impact factor $\tit{Impact}[i, l_\tit{cand}]$ is computed by the AND of the two components, \textit{i.e.}, $(n[i] \oplus n'[i]) \land P[i, n, \hat y_k]$, as shown in Eq.~\eqref{eq:po_update}.
Finally, the new PO value $y'_k[i]$ is obtained by XORing the original PO value $\hat y_k[i]$ with the binary impact factor $\tit{Impact}[i, l_\tit{cand}]$.

For each candidate LAC $l_\tit{cand}$,
after obtaining the new PO values $y'_k[i]$'s for all POs under all simulation patterns,
we can further obtain the deviation between the PO values of the accurate circuit $G$
and 
those of the approximate circuit after applying $l_\tit{cand}$.
Then, $\tit{MaxError}_\tit{LB}(l_\tit{cand})$ can be obtained as the maximum deviation over all simulation patterns.

\begin{example}
For the example circuit in Fig.~\ref{fig:cpm}, 
consider a LAC $l_\tit{cand}$ that replaces node $n$ with another node $x_2$.
Before applying $l_\tit{cand}$,
the values of $\hat y_0$ and $\hat y_1$ under the $i$-th simulation pattern,
denoted as $\hat y_0[i]$ and $\hat y_1[i]$, are 0 and 1, respectively.
From Example~\ref{ex:cpm},
we know $P[i, n, \hat y_0] = 1$ and $P[i, n, \hat y_1] = 0$.
Then, the impact factors can be computed as follows:
\begin{equation*}
\begin{split}
    \tit{Impact}[i, l_\tit{cand}, \hat y_0] & = (n[i] \oplus x_2[i]) \land P[i, n, \hat y_0] = 1, \\
    \tit{Impact}[i, l_\tit{cand}, \hat y_1] & = (n[i] \oplus x_2[i]) \land P[i, n, \hat y_1] = 0.
\end{split}
\end{equation*}
This means that applying $l_\tit{cand}$ changes $\hat y_0$'s value under the $i$-th pattern,
while it does not affect $\hat y_1$'s value.
Therefore, after applying $l_\tit{cand}$,
the new values of $\hat y_0$ and $\hat y_1$ under the $i$-th simulation pattern,
denoted as $y'_0[i]$ and $y'_1[i]$, can be updated as follows:
\begin{equation*}
\begin{split}
    y'_0[i] & = \hat y_0[i] \oplus \tit{Impact}[i, l_\tit{cand}, \hat y_0] = 0 \oplus 1 = 1, \\
    y'_1[i] & = \hat y_1[i] \oplus \tit{Impact}[i, l_\tit{cand}, \hat y_1] = 1 \oplus 0 = 1.
\end{split}
\end{equation*}

If we consider the MaxED metric
and assume that the PO values of the accurate circuit $G$ under the $i$-th simulation pattern are $y_0[i] = 0$ and $y_1[i] = 1$,
then the deviation between the PO values of $G$ and those of the approximate circuit after applying $l_\tit{cand}$ under the $i$-th simulation pattern is
\begin{equation*}
    D\left(\bm y[i], \bm{\hat y}[i]\right) 
    = \left|(2 y_1[i] + y_0[i]) - (2 y'_1[i] + y'_0[i])\right| = 1. \\
\end{equation*}%
After computing the deviation $D\left(\bm y[i], \bm{\hat y}[i]\right)$ for all simulation patterns,
the maximum value of the deviation is $\tit{MaxError}_\tit{LB}(l_\tit{cand})$, \ie{}, the lower bound on the maximum error caused by the LAC $l_\tit{cand}$.
\end{example}

Using the CPM-based method to compute the lower bounds on the maximum errors of all candidate LACs,
the main computational effort lies in constructing the CPM.
As mentioned above,
obtaining the CPM for the current approximate circuit $\hat G$ requires only $O(N)$ simulation runs,
where $N$ is the number of functional nodes in $\hat G$.
Compared with the naive method with $O(|L_\tit{cand}|)$ simulation runs,
the CPM-based method is much more efficient,
as $N$ is usually much smaller than $|L_\tit{cand}|$.
After obtaining the lower bounds on the maximum errors of all candidate LACs,
the LAC pruning step can efficiently filter out the invalid LACs that violate the maximum error constraint based on Eq.~\eqref{eq:maxerr_lb}
and return the set of remaining LACs $L_\tit{rem}$ to be further checked by the promising LAC selection and application step.

\subsection{Promising LAC Selection and Application Based on Simulation-Guided SAT Solving}\label{subsec:sat}

As shown in Fig.~\ref{fig:ALS_flow},
the promising LAC selection and application step is responsible for checking the validity of the remaining LACs in $L_\tit{rem}$ after the simulation-based pruning,
selecting a subset of promising LACs $L_\tit{prom}$ to reduce the circuit area as much as possible,
and applying all LACs in $L_\tit{prom}$ to simplify the current approximate circuit $\hat G$.

In the following parts,
we will first formulate the LAC selection problem in Section~\ref{subsubsec:problem}
and then introduce a greedy LAC selection strategy supported by simulation-guided SAT solving in Section~\ref{subsubsec:sat_select}.
Finally, we will discuss the order of checking LACs in Section~\ref{subsec:sorter},
which significantly affects the quality of the final approximate circuit and
the efficiency of the promising LAC selection and application step.

\subsubsection{Formulation of the LAC Selection Problem}\label{subsubsec:problem}
The LAC selection problem can be formulated as: Given a current approximate circuit $\hat G$ and a set of LACs $L_\tit{rem}$,
find a subset of promising LACs $L_\tit{prom} \subseteq L_\tit{rem}$ to reduce the circuit area as much as possible,
while satisfying the following three constraints:
\begin{itemize}
    \item Error constraint: the maximum error caused by applying all LACs in $L_\tit{prom}$ is within the error bound $B$. 
    \item Circuit integrity constraint: the LACs in $L_\tit{prom}$ do not introduce a logic loop in the circuit.
    \item LAC conflict constraint: at most one LAC can be applied to each node in the circuit, since two LACs cannot be applied to the same node simultaneously.
\end{itemize}%
\radd{Note that our framework supports circuits represented in any graph format,
such as AIGs and gate netlists.
For AIGs, circuit area is estimated by counting the number of AND nodes,
while for gate netlists, it is obtained by summing the areas of all gates.}

Our LAC selection problem is similar to the one in the MUSCAT method~\cite{witschen2022muscat},
which is a state-of-the-art ALS method under the maximum error constraint.
However, the formulation of MUSCAT only considers the constant LACs that replace a signal by a constant 0 or 1,
while ours considers arbitrary single-output LACs.
To solve this NP-hard combinatorial optimization problem,
MUSCAT converts the problem into a MUS problem and solves it using MUS solvers,
which is time-consuming and limits its scalability.
Although MUSCAT sets a time limit for the MUS solving to enhance the scalability,
the time limit leads to suboptimal approximate circuits.
In our work, we propose a greedy selection strategy to obtain a good solution efficiently,
which will be introduced next.

\subsubsection{Greedy LAC Selection Strategy Supported by Simulation-Guided SAT Solving}\label{subsubsec:sat_select}

\begin{figure}[!htbp]
    \centering
    \includegraphics[width=1.0\columnwidth]{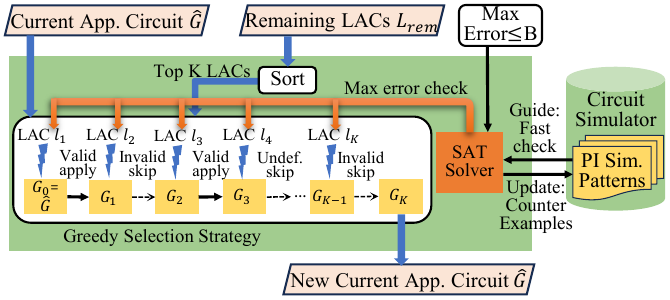}
    \vspace{-2.0em}
    \caption{Greedy promising LAC selection strategy supported by simulation-guided SAT solving.}
    \label{fig:lac_select}
\end{figure}

As shown in Fig.~\ref{fig:lac_select}, 
our method first sorts the LACs in $L_\tit{rem}$ and keeps the top $K$,
where $K$ is a user-defined parameter.
More details about the sorting are discussed in Section~\ref{subsec:sorter}.
Then, each LAC in the sorted list is examined in order to determine whether it should be applied.
Denote the initial circuit before applying any LAC as $G_0 \!=\! \hat G$ (the current approximate circuit)
and the updated circuit after processing the $j$-th ($1\!\le\! j\!\le\! K$) LAC as $G_j$.
After processing all $K$ LACs, the resulting circuit $G_K$ becomes the updated current approximate circuit $\hat G$ for subsequent iterations.
Note that the $j$-th candidate LAC $l_j$ affects the circuit $G_{j-1}$.
When we process LAC $l_j$,
we first check whether applying it to the circuit $G_{j-1}$ will introduce a logic loop in the circuit.
If so, we skip $l_j$ and keep the circuit $G_j$ the same as $G_{j-1}$.
Then, for the LAC $l_j$ that does not introduce a logic loop,
we check its validity using the error miter-based method introduced in Section~\ref{subsec:approx_miter}.
Specifically,
we build an error miter (see Fig.~\ref{fig:error_miter}) using the accurate circuit $G$ and the approximate circuit after applying $l_j$ to $G_{j-1}$,
convert the miter into a SAT problem,
and then use a SAT solver to solve the problem.
The solution of the SAT problem determines whether to apply $l_j$ or not.
There are three possible results of the SAT solving for $l_j$:

\begin{itemize}

\item UNSAT: This indicates that the maximum error caused by LAC $l_j$ is no larger than $B$.
In this case, LAC $l_j$ is valid, and we apply it to the circuit $G_{j-1}$, obtaining the resulting circuit $G_j$.
Examples of this case are $l_1$ and $l_3$ in Fig.~\ref{fig:lac_select}.
Note that in order to satisfy the LAC conflict constraint,
after applying $l_j$,
we remove all LACs that affect the same node as $l_j$ from \cdel{$L_\tit{rem}$}the top $K$ LACs.

\item SAT: This indicates that applying LAC $l_j$ to the circuit $G_{j-1}$ causes a maximum error that exceeds $B$.
In this case, LAC $l_j$ is invalid and should be skipped,
and circuit $G_j$ keeps the same as circuit $G_{j-1}$.
Examples of this case are $l_2$ and $l_K$ in Fig.~\ref{fig:lac_select}.

\item UNDEFINED: This happens when the SAT solver cannot give a solution within a computing resource limit.
In this case, the solver cannot determine whether $l_j$ is valid or not.
To avoid violating the error constraint,
conservatively, we do not select $l_j$ and skip it,
and circuit $G_j$ keeps the same as circuit $G_{j-1}$.
In our implementation, we set a maximum conflict number for the SAT solver to avoid the long runtime of SAT solving.
In a SAT solver, a conflict happens when the current variable assignments make a clause false.
During SAT solving, if the number of conflicts exceeds the maximum conflict number, the solver returns UNDEFINED.
Moreover, if the SAT solver returns UNDEFINED for LAC $l_j$,
for efficiency,
we add $l_j$ into a blacklist and do not consider it again in future iterations of the ALS flow.
An example of this case is $l_4$ in Fig.~\ref{fig:lac_select}.
\end{itemize}

\vspace{-1.0em}

When the solver returns SAT for LAC $l_j$,
a counter-example pattern is also returned,
which is a PI pattern that causes the maximum error to exceed $B$.
Note that the counter-example pattern is not sampled in previous logic simulation in the LAC pruning step.
It can be viewed as a sensitive pattern that activates the large deviation caused by LAC $l_j$.
This sensitive pattern is very likely to activate large deviations caused by other LACs,
especially those affecting the same node as $l_j$.
Therefore, the counter-example pattern is very helpful in guiding the future maximum error checking of other LACs.

We propose a reuse mechanism of the counter-example patterns,
as shown in the right part of Fig.~\ref{fig:lac_select}.
\radd{When examining LAC $l_j$,
before solving the SAT problem for $l_j$,
we perform logic simulation using the counter-example patterns generated by $l_s$'s ($s < j$) \rdel{to quickly check the maximum error caused by $l_j$}to quickly check whether $l_j$ already violates the maximum error bound under these patterns}.
If simulation finds that $l_j$ is invalid,
then we directly skip the LAC without solving the SAT problem.
Otherwise, we still need to solve the SAT problem for $l_j$ to check its validity.
By guiding the SAT solving with the simulation using the counter-example patterns,
we can skip unnecessary SAT solving and improve the efficiency of the LAC selection and application step.
Furthermore, we also update the simulation patterns by storing the counter-example patterns.
In this way, the stored counter-example patterns will guide the future iterations of the ALS flow in their simulation-based LAC pruning and promising LAC selection and application steps.
In practice, this method reduces the number of SAT problems to be solved in the ALS flow and accelerates the ALS flow.

\subsubsection{The Order of Checking and Selecting LACs}\label{subsec:sorter}

As shown in Fig.~\ref{fig:lac_select},
before applying the greedy LAC selection strategy,
an essential step is to sort the LACs in $L_\tit{rem}$.
After sorting, the top $K$ LACs are kept and the other LACs are discarded.

The sorting of LACs is crucial for both the quality of the final approximate circuit and the efficiency of the promising LAC selection and application step.
Note that in the greedy LAC selection strategy,
a previously selected LAC changes the circuit structure,
and hence affects the validity of the following LACs.
For example, in Fig.~\ref{fig:lac_select},
$l_1$ is selected and applied to the circuit,
affecting the validity of $l_2, l_3, \ldots, l_K$.
Thus, different orders of LACs may lead to different results of the LAC selection
and finally affect the quality of the final approximate circuit.
Moreover, a poor order of LACs may slow down the LAC selection process.
For example, if invalid LACs are checked first,
then the promising LAC selection and application step may spend a lot of time checking and discarding them
with no simplification of the circuit, hence wasting time.

\radd{We propose a sorting strategy based on two criteria: 
1) the maximum error lower bound (primary) and 
2) the estimated area reduction of the LAC (secondary).}
\radd{First, the LACs are sorted in the ascending order of their maximum error lower bounds,
obtained from logic simulation.}
The motivation is that only the top $K$ LACs are kept after the sorting.
\rdel{and we want to keep the valid LACs in the top $K$ LACs as many as possible}\radd{We want to keep as many valid LACs as possible within the top $K$}
so that the circuit area can be effectively reduced by applying these valid LACs.
Given that a LAC with a small maximum error lower bound is more likely to be valid,
sorting the LACs by the lower bounds can increase the probability of keeping valid LACs in the top $K$ LACs.

\radd{Only if multiple LACs share the same maximum error lower bound,
they are further sorted in descending order of the estimated area reduction,
prioritizing LACs that reduce more area.
Our method supports circuits represented in any graph format, such as AIGs and gate netlists.
For AIGs,
a LAC's area reduction is estimated by counting the nodes in the maximum fanout-free cone (MFFC)~\cite{cong1993area} of the node where the LAC is applied,
instead of performing time-consuming logic synthesis to obtain the exact reduction.
For gate netlists,
a LAC's area reduction is obtained by summing the areas of the gates in the corresponding MFFC.}
\section{Experimental Results}\label{sec:res}

\subsection{Experimental Setup}\label{subsec:res-setup}

We implement the proposed ALS flow under the maximum error constraint in C++ and test it on a single core of an AMD Ryzen 9 5900X processor with 64GB RAM.
Our flow is developed upon a state-of-the-art open-source logic synthesis and verification system, ABC~\cite{berkeley2024abc}.
The flow also integrates CryptoMiniSat~\cite{soos2009extending},
a modern SAT solver with rich features and a simple interface.
To avoid long runtime of SAT solving,
we set a maximum conflict number of $2^{18}$ for each SAT problem.

In all experiments,
the initial circuits are represented as AIGs,
although our flow can also handle other circuit representations, such as gate netlists.
\cdel{the original circuits are first converted into AIGs and then simplified by our flow.}%
The reason for using AIGs is that many AIG-based logic synthesis works~\cite{riener2019scalable,calvino2022versatile} and ALS works~\cite{meng2023hedals,lee2023approximate} have shown significant advantages in reducing hardware cost, particularly for CMOS technologies.
The standard cell library used in our experiments is the Nangate 45nm library~\cite{nangate2022nangate}.
\radd{The traditional logic synthesis in the last step of our flow (see Fig.~\ref{fig:ALS_flow})
consists of AIG optimization and technology mapping.
For AIG optimization, we apply the ABC script \tit{``resyn2rs''} for 3 iterations.
This script combines four transformations, \ie{}, 
\tit{rewriting}, \textit{refactoring}, \textit{balancing}, and \textit{resubstitution}, 
to simplify the local structure of the AIG,
and has been shown effective and widely used in recent works~\cite{sun2024massively,costamagna2025area,calvino2024enhancing,testa2020extending}.
Iteratively applying the script can further reduce the AIG size and depth,
and in our experiments, 
3 iterations provide a good trade-off between circuit quality and runtime.
For technology mapping, we use the standard area-oriented mapping script \tit{``dch; amap''} to convert the AIG into a gate netlist,
as recommended in the ABC tutorial~\cite{mishchenko2019introduction}.}
Unless otherwise specified, 
the following default parameters are used in all experiments.
For the simulation-based LAC pruning (see Section~\ref{subsubsec:prune_theory}),
we first use $M_\tit{small} = 2^{10}$ simulation patterns to quickly filter out LACs inducing large errors,
and then use $M=2^{13}$ simulation patterns for a more fine-grained filtering.
For the parameter $K$ in the sorting strategy in Section~\ref{subsec:sorter},
we set $K=100$\cdel{ in all experiments}.
To build the error miter in Fig.~\ref{fig:error_miter} for maximum error checking,
we use Verilog to describe the error miter
and then use Yosys~\cite{yosys2025yosys} and ABC to synthesize the Verilog description.

\radd{To evaluate the hardware cost}\rdel{ of a circuit},
\radd{we measure the area, delay, and power consumption of the synthesized gate netlist (post-synthesis, before place-and-route).
The area is computed by summing the individual gate areas in the netlist.
The delay is obtained using the static timing analysis command \tit{``stime''} in ABC.
The power is estimated with the Synopsys Design Compiler~\cite{synopsys2025synopsys} at 2MHz under a uniform input distribution.}
\radd{We then define \tit{area ratio}, \tit{delay ratio}, and \tit{power ratio} as the respective values of the approximate netlist normalized to those of the accurate one.}
Smaller ratios are preferred.
To evaluate the accuracy of circuits,
two different maximum error metrics, MaxED and MaxHD, are considered in our experiments.
Note that the focus of this work is ALS under the maximum error constraint,
so we do not compare it with other ALS methods under \fvdel{the }average error constraints, such as~\cite{lee2023approximate} and~\cite{ma2021approximate}.
\rdel{For all generated the approximate circuits}\radd{For all generated approximate circuits},
the error miter in Fig.~\ref{fig:error_miter} is used to formally verify that the maximum errors of the circuits satisfy the given error bounds.

\begin{table}[!htbp]
\centering
\caption{Experimental benchmarks. \rdel{Area and delay}\radd{Area, delay, and power} are measured by mapping AIGs\cadd{ into gate netlists} with the Nangate 45nm library.}\label{tab:benchmark}
\vspace{-1em}
\begin{threeparttable}
\tabcolsep=2.0pt

\begin{tabular}{clrrrrrr}
\toprule
\multirow{2}{*}{\makecell{Benchmark\\suite}} & \multirow{2}{*}{Circuit} & \multirow{2}{*}{\#PIs/\#POs} & \multicolumn{2}{c}{AIG} & \multicolumn{3}{c}{Gate netlist} \\
                                             &         &             & Size       & Depth      & \makecell{Area\\/$um^2$}  & \makecell{Delay\\/$ns$}  & \radd{\makecell{Power\\/$\mu W$}}\\ \midrule
\multirow{9}{*}{\makecell{Used in\\MECALS\\\cite{meng2023mecals}\cdel{\\(arithmetic)}}}
                                             & absdiff & 16/8        & 141        & 14         & 87.3      & 0.42      & \radd{80.9}  \\
                                             & add8    & 16/9        & 66         & 10         & 42.0      & 0.36      & \radd{36.8}  \\
                                             & add32   & 64/33       & 252        & 64         & 184.6     & 1.84      & \radd{173.0}  \\
                                             & binsqrd & 16/18       & 1562       & 50         & 1052.3    & 1.53      & \radd{989.5}  \\
                                             & buttfly & 32/34       & 265        & 48         & 170.5     & 1.01      & \radd{172.6}  \\
                                             & mac     & 12/8        & 145        & 20         & 92.8      & 0.60      & \radd{73.4}  \\
                                             & mult8   & 16/16       & 649        & 40         & 435.4     & 1.26      & \radd{410.2}  \\
                                             & mult16  & 32/32       & 1981       & 72         & 1418.8    & 1.98      & \radd{1707.2}  \\
                                             & mult32  & 64/64       & 8340       & 53         & 5723.3    & 1.87      & \radd{7868.3}  \\ \midrule
\multirow{8}{*}{\makecell{EPFL\tnote{*}\\arith-\\metic}}    & add128  & 256/129     & 1297       & 28         & 933.4     & 0.96      & \radd{825.0}  \\
                                             & bar     & 135/128     & 2688       & 14         & 1267.8    & 0.92      & \radd{1753.1}  \\
                                             & log2    & 32/32       & 38540      & 419        & 21480.6   & 14.26     & \radd{40410.0}  \\
                                             & max     & 512/130     & 2686       & 549        & 1646.3    & 15.81     & \radd{1971.2}  \\
                                             & mult64  & 128/128     & 33242      & 326        & 16447.3   & 9.47      & \radd{31105.0}  \\
                                             & sine    & 24/25       & 7044       & 180        & 4112.1    & 5.93      & \radd{5968.1}  \\
                                             & sqrt    & 128/64      & 21951      & 4591       & 13464.1   & 216.92    & \radd{58397.0}  \\
                                             & square  & 64/128      & 20030      & 296        & 12801.8   & 7.96      & \radd{18836.0}  \\ \midrule
\multirow{7}{*}{\makecell{ISCAS85}}        & c880    & 60/26       & 313        & 22         & 198.2     & 0.59      & \radd{129.8}  \\
                                             & c1355   & 41/32       & 390        & 16         & 235.9     & 0.56      & \radd{260.1}  \\
                                             & c1908   & 33/25       & 367        & 25         & 229.6     & 0.86      & \radd{222.6}  \\
                                             & c2670   & 233/140     & 579        & 17         & 385.2     & 0.68      & \radd{325.4}  \\
                                             & c3540   & 50/22       & 937        & 32         & 521.1     & 1.02      & \radd{404.4}  \\
                                             & c5315   & 178/123     & 1306       & 28         & 720.3     & 0.72      & \radd{643.7}  \\
                                             & c7552   & 207/108     & 1469       & 26         & 903.6     & 1.43      & \radd{898.8}  \\ \bottomrule
\end{tabular}

\begin{tablenotes}
    \item[*] The large benchmark \tit{hyp} is omitted and cannot be handled by both our and baseline methods.
    The \tit{div} benchmark is omitted since there is no space of approximation under the given MaxED bounds.
\end{tablenotes}
\end{threeparttable}
\end{table}

\radd{The benchmarks used in our experiments are listed in Table~\ref{tab:benchmark},
which includes circuit names, PI/PO numbers, AIG size, AIG depth, circuit area, circuit delay, and circuit power}.
They are the benchmarks used in MECALS~\cite{meng2023mecals},
EPFL arithmetic benchmarks~\cite{lsi2024epfl}, 
and ISCAS85 benchmarks~\cite{hansen1999unveiling}.
The initial AIGs have been well optimized to ensure as little redundancy as possible. 
These AIGs are then used as input to our ALS flow and those ALS flows for comparison.
The baseline methods are the MECALS~\cite{meng2023mecals} and MUSCAT~\cite{witschen2022muscat} methods.
MECALS is a state-of-the-art ALS method,
in which the maximum error checking problem is converted into a SAT sweeping problem.
MUSCAT is another state-of-the-art method that converts the ALS problem under maximum error constraint into a MUS problem and solves it using a MUS solver.
\radd{MECALS can handle both constant~\cite{shin2011new} and SASIMI~\cite{venkataramani2013substitute} LACs,
while MUSCAT only supports constant LACs.
In our experiments, 
for a circuit with $N$ nodes,
our method and MECALS consider $2N$ constant LACs
and $O(N^2)$ SASIMI LACs,
while MUSCAT only considers the $2N$ constant LACs.}

\begin{table*}[!htbp]
\caption{Comparison of our method with the state-of-the-art methods under the MaxED constraint.
\radd{\tbf{Bold} entries indicate the smallest area, delay, and power ratios, or the shortest runtime.}
N/A means MUSCAT cannot obtain \cdel{the }final approximate circuits in 24 hours.}\label{tab:res_maxed}
\vspace{-1em}
\centering
\tabcolsep=3pt

\begin{tabular}{cr|rrr|rrr|rrr|rrr}
\bottomrule
\multirow{2}{*}{Circuit} & \multirow{2}{*}{\makecell{MaxED\\bound}} & \multicolumn{3}{c|}{Area ratio}             & \multicolumn{3}{c|}{Delay ratio}                & \multicolumn{3}{c|}{\radd{Power ratio}}                     & \multicolumn{3}{c}{Runtime/s}                \\
                         &             & Ours            & MECALS          & MUSCAT & Ours             & MECALS          & MUSCAT          & \radd{Ours}            & \radd{MECALS}          & \radd{MUSCAT}          & Ours          & MECALS        & MUSCAT       \\ \hline
\multirow{2}{*}{absdiff} & 1           & \textbf{64.0\%} & 65.2\%          & 86.2\% & \textbf{107.2\%} & 108.9\%         & 94.6\%          & \radd{68.1\%}          & \radd{\textbf{65.2\%}} & \radd{83.7\%}          & \textbf{0.2}  & 0.4           & 1.1          \\
                         & 3           & \textbf{59.7\%} & 69.2\%          & 84.7\% & \textbf{95.1\%}  & 106.2\%         & 105.7\%         & \radd{\textbf{59.5\%}} & \radd{62.4\%}          & \radd{70.3\%}          & 1.7           & \textbf{0.6}  & 1.1          \\ \hline
\multirow{2}{*}{add8}    & 1           & \textbf{76.0\%} & 92.5\%          & 95.0\% & \textbf{77.3\%}  & 88.1\%          & 101.6\%         & \radd{\textbf{81.4\%}} & \radd{92.3\%}          & \radd{94.3\%}          & 0.12          & \textbf{0.11} & 0.7          \\
                         & 3           & \textbf{72.2\%} & 81.7\%          & 85.5\% & \textbf{76.7\%}  & 95.7\%          & 87.0\%          & \radd{\textbf{76.7\%}} & \radd{80.7\%}          & \radd{85.6\%}          & 0.7           & \textbf{0.2}  & 1.0          \\ \hline
\multirow{2}{*}{add32}   & 9           & \textbf{70.3\%} & 71.9\%          & 84.1\% & 95.0\%           & 99.6\%          & \textbf{76.3\%} & \radd{\textbf{77.6\%}} & \radd{79.6\%}          & \radd{100.0\%}         & \textbf{28}   & 50            & 23           \\
                         & 97          & 63.0\%          & \textbf{62.5\%} & 82.6\% & 84.9\%           & 86.0\%          & \textbf{83.7\%} & \radd{69.5\%}          & \radd{\textbf{68.5\%}} & \radd{82.6\%}          & 114           & 84            & \textbf{28}  \\ \hline
\multirow{2}{*}{binsqrd} & 3           & \textbf{78.4\%} & 79.9\%          & 99.9\% & 95.2\%           & \textbf{93.8\%} & 100.0\%         & \radd{\textbf{83.1\%}} & \radd{84.0\%}          & \radd{100.4\%}         & \textbf{15}   & 1432          & 377          \\
                         & 12          & \textbf{76.9\%} & 78.6\%          & 97.2\% & 95.2\%           & \textbf{93.9\%} & 99.7\%          & \radd{\textbf{81.6\%}} & \radd{82.7\%}          & \radd{98.2\%}          & \textbf{271}  & 8947          & 7230         \\ \hline
\multirow{2}{*}{buttfly} & 10          & \textbf{79.4\%} & 86.9\%          & 97.7\% & \textbf{81.6\%}  & 91.8\%          & 99.3\%          & \radd{\textbf{81.5\%}} & \radd{83.6\%}          & \radd{93.1\%}          & 15            & 3.6           & \textbf{2.2} \\
                         & 111         & \textbf{74.6\%} & 85.3\%          & 94.9\% & 101.1\%          & \textbf{83.7\%} & 99.3\%          & \radd{\textbf{77.3\%}} & \radd{78.8\%}          & \radd{86.9\%}          & 56            & 5.8           & \textbf{2.3} \\ \hline
\multirow{2}{*}{mac}     & 1           & 87.1\%          & \textbf{86.6\%} & 96.6\% & \textbf{94.4\%}  & 95.8\%          & 99.6\%          & \radd{92.7\%}          & \radd{\textbf{91.3\%}} & \radd{92.4\%}          & \textbf{0.3}  & 1.0           & 1.9          \\
                         & 2           & 85.1\%          & \textbf{82.8\%} & 89.4\% & 94.4\%           & 91.5\%          & \textbf{87.4\%} & \radd{88.3\%}          & \radd{87.8\%}          & \radd{\textbf{82.4\%}} & \textbf{0.8}  & 4.5           & 4.5          \\ \hline
\multirow{2}{*}{mult8}   & 3           & \textbf{74.1\%} & 75.1\%          & 98.2\% & \textbf{81.1\%}  & 82.0\%          & 100.0\%         & \radd{75.1\%}          & \radd{\textbf{72.4\%}} & \radd{98.6\%}          & \textbf{1.3}  & 78            & 949          \\
                         & 9           & \textbf{72.5\%} & 73.1\%          & 96.3\% & 85.9\%           & \textbf{81.9\%} & 100.0\%         & \radd{76.3\%}          & \radd{\textbf{70.8\%}} & \radd{97.3\%}          & \textbf{7.7}  & 111           & 7225         \\ \hline
\multicolumn{2}{c|}{Average of above}   & \textbf{73.8\%} & 77.9\%          & 92.0\% & \textbf{90.4\%} & 92.8\%          & 95.3\%          & \radd{\textbf{77.8\%}} & \radd{78.6\%}          & \radd{90.4\%}          & \textbf{37}   & 766           & 1132         \\ \hline \hline
\multirow{2}{*}{mult16}  & 9           & \textbf{98.3\%} & 99.7\%          & N/A    & \textbf{98.1\%}  & 100.7\%         & N/A             & \radd{\textbf{99.0\%}} & \radd{99.9\%}          & \radd{N/A}             & \textbf{9.3}  & 154           & N/A          \\
                         & 84          & \textbf{93.7\%} & 95.7\%          & N/A    & \textbf{92.1\%}  & 97.1\%          & N/A             & \radd{\textbf{94.5\%}} & \radd{96.0\%}          & \radd{N/A}             & \textbf{83}   & 368           & N/A          \\ \hline
\multirow{2}{*}{mult32}  & 84          & \textbf{98.0\%} & 99.6\%          & N/A    & \textbf{97.5\%}  & 102.0\%         & N/A             & \radd{\textbf{98.2\%}} & \radd{99.1\%}          & \radd{N/A}             & \textbf{203}  & 997           & N/A          \\
                         & 7131        & \textbf{93.9\%} & 98.6\%          & N/A    & 100.9\%          & \textbf{93.8\%} & N/A             & \radd{\textbf{94.7\%}} & \radd{97.8\%}          & \radd{N/A}             & \textbf{1544} & 3364          & N/A          \\ \hline
\multicolumn{2}{c|}{Average of all}     & \textbf{78.7\%} & 82.5\%          & N/A    & \textbf{91.9\%} & 94.0\%          & N/A             & \radd{\textbf{82.0\%}} & \radd{82.9\%}          & \radd{N/A}             & \textbf{131}  & 867           & N/A          \\ \toprule
\end{tabular}
\vspace{-2em}
\end{table*}

\subsection{Experiments Under the MaxED Constraint}\label{subsec:res-maxed}
This set of experiments tests the arithmetic benchmarks used in MECALS and from the EPFL benchmark suite in Table~\ref{tab:benchmark} under the MaxED constraint.
Note that MaxED is a suitable error metric for arithmetic circuits,
because from Eq.~\eqref{eq:dev_maxed},
the deviation function of MaxED considers the different significance of different POs,
which measures the absolute difference between the numerical values encoded by the POs of accurate and approximate circuits.
In what follows, we first compare our flow with the state-of-the-art methods on the benchmarks used in MECALS
and then show the scalability of our flow using the EPFL arithmetic benchmarks.

\subsubsection{Comparison with State-of-the-Art Methods}

We compare our ALS flow with MECALS and MUSCAT.
The tested benchmarks are those used in MECALS.
We run the open-source codes of MECALS and MUSCAT on the same platform for fair comparison.
The MaxED bounds are set to \rdel{$\lfloor 2^{0.1O} \rfloor$ and $\lfloor 2^{0.2O} \rfloor$}\radd{$\lfloor 2^{O/10} \rfloor$ and $\lfloor 2^{O/5} \rfloor$} for each benchmark,
where $O$ is the number of POs of the benchmark, 
and the function $\lfloor x \rfloor$ gives the greatest integer less than or equal to $x$.
In each iteration of our flow,
the LAC generation step (see Fig.~\ref{fig:ALS_flow}) produces both constant and SASIMI LACs.

Table~\ref{tab:res_maxed} compares the area ratio, delay ratio,\radd{ power ratio,} and runtime of our flow, MECALS, and MUSCAT under the MaxED constraint.
\rdel{The bold entries indicate that our method outperforms both MECALS and MUSCAT,}%
\radd{The bold entries indicate the smallest area ratio, delay ratio, power ratio, and runtime among the three methods,}
and we use the same highlighting in the following tables.
We can see that our method achieves the smallest area ratios for most benchmarks.
\radd{For the first 7 smaller benchmarks, 
our method achieves an average area, delay, and power ratio of $73.8\%$, $90.4\%$, and $77.8\%$, respectively.
Compared to MECALS, our method reduces area, delay, and power by $4.1\%$, $2.4\%$\, and 0.8\%\cdel{, respectively,} on average.
Compared to MUSCAT, our method reduces area, delay, and power by 18.2\%, 4.9\%, and 12.6\%\cdel{, respectively,} on average.}
Moreover, MUSCAT cannot obtain the final approximate circuit in 24 hours for the benchmarks \tit{mult16} and \tit{mult32},
while our method and MECALS can.
\radd{Over all benchmarks, our method reduces area, delay, and power by $3.8\%$, $2.1\%$, and 0.9\%\cdel{,} on average compared to MECALS.
Note that for the benchmarks \tit{add8} and \tit{mult16},
our method achieves smaller area, delay, and power ratios than MECALS for both MaxED bounds.}
However, for the benchmark \tit{add32} under the MaxED bound of $97$ and the benchmark \tit{mac},
our method is worse than MECALS in terms of the area ratio.
One possible reason is as follows.
Both our flow and MECALS iteratively simplify the circuit.
In each iteration, our flow selects multiple promising LACs (see Section~\ref{subsubsec:sat_select}),
while MECALS only selects one.
This difference may lead to better performance of MECALS on some benchmarks.
However, our flow reduces more area than MUSCAT for all benchmarks.
This is because our flow can handle more complex LACs than MUSCAT,
which can achieve better approximate circuits.

Our method is more efficient than MECALS and MUSCAT.
Over the first 7 benchmarks, our method is $20.7\times$ faster than MECALS and $30.6\times$ faster than MUSCAT on average.
Over all benchmarks, our method speeds up by $6.6\times$ on average compared to MECALS.
Note that our method is slower than MECALS for some benchmarks.
For the small benchmarks \tit{absdiff} and \tit{add8},
the total runtime of our method is within $2$ seconds,
so the runtime difference is negligible.
For the benchmark \tit{buttfly}, compared to MECALS,
our method takes more time but dramatically reduces the area,
which is worth the trade-off.
For the benchmark \tit{add32} under the MaxED bound of $97$,
our method consumes more time than MECALS and MUSCAT,
while the area and delay ratios are still competitive.
Additionally,
across all benchmarks in Table~\ref{tab:res_maxed},
our method finishes after an average of 5.8 iterations.
The average number of LACs applied per iteration is 7.4,
obtained by dividing the total number of applied LACs across all iterations and benchmarks over the total number of iterations across all benchmarks.
In contrast, MECALS needs an average of 8.3 iterations
with only one LAC applied per iteration.
Applying multiple LACs per iteration reduces the number of iterations, 
thereby shortening the overall runtime.

\begin{table*}[!htbp]
\caption{Comparison of our ALS flow with the MECALS method on the EPFL arithmetic benchmarks under the MaxED constraint.
\radd{\tbf{Bold} entries indicate smaller area, delay, and power ratios, or shorter runtime.}
N/A means that MECALS cannot obtain the final approximate circuit in 24 hours.}\label{tab:res_maxed_epfl}
\vspace{-1em}
\centering

\begin{tabular}{cr|rr|rr|rr|rr}
\bottomrule
\multirow{2}{*}{Circuit} & \multirow{2}{*}{\makecell{MaxED\\bound}} & \multicolumn{2}{c|}{Area ratio} & \multicolumn{2}{c|}{Delay ratio} & \multicolumn{2}{c|}{\radd{Power ratio}} & \multicolumn{2}{c}{Runtime/s} \\
                         &                                          & Ours               & MECALS    & Ours                & MECALS    & \radd{Ours}               & \radd{MECALS}    & Ours              & MECALS    \\ \hline
\multirow{2}{*}{add128}  & 7.6E+03                                  & \textbf{86.9\%}    & 94.2\%    & 105.4\%             & \textbf{96.4\%}    & \radd{\textbf{86.6\%}}    & \radd{92.0\%}    & \textbf{7.2}      & 474       \\
                         & 5.8E+07                                  & \textbf{77.6\%}    & 84.1\%    & \textbf{99.9\%}     & 100.1\%   & \radd{\textbf{77.1\%}}    & \radd{81.2\%}    & \textbf{11}       & 3264      \\ \hline
\multirow{2}{*}{bar}     & 7.1E+03                                  & \textbf{97.5\%}    & 97.9\%    & \textbf{99.9\%}     & 100.6\%   & \radd{\textbf{95.1\%}}    & \radd{96.0\%}    & \textbf{2.2}      & 412       \\
                         & 5.1E+07                                  & 96.2\%             & \textbf{95.7}\%    & \textbf{100.2\%}    & 100.4\%   & \radd{90.8\%}             & \radd{\textbf{90.3\%}}    & \textbf{2.6}      & 753       \\ \hline
\multirow{2}{*}{max}     & 8.2E+03                                  & 94.2\%             & 94.2\%    & 82.4\%              & 82.4\%    & \radd{90.5\%}             & \radd{\textbf{90.1\%}}    & \textbf{273}      & 6362      \\
                         & 6.7E+07                                  & \textbf{93.1\%}    & 93.2\%    & \textbf{80.8\%}     & 80.9\%    & \radd{\textbf{87.5\%}}    & \radd{87.9\%}    & \textbf{303}      & 8580      \\ \hline
\multirow{2}{*}{mult64}  & 7.1E+03                                  & \textbf{96.1\%}    & 99.1\%    & 100.6\%             & \textbf{98.5\%}    & \radd{\textbf{98.3\%}}    & \radd{101.0\%}   & \textbf{4347}     & 45437     \\
                         & 5.1E+07                                  & \textbf{95.7\%}    & 98.2\%    & 102.4\%             & \textbf{99.2\%}    & \radd{\textbf{97.8\%}}    & \radd{99.6\%}    & \textbf{1996}     & 78970     \\ \hline
square                   & 7.1E+03                                  & \textbf{92.9\%}    & 99.4\%    & \textbf{93.3\%}     & 98.9\%    & \radd{\textbf{96.3\%}}    & \radd{99.1\%}    & \textbf{1504}     & 53925     \\ \hline
\multicolumn{2}{c|}{Average of above}                                & \textbf{92.2\%}    & 95.1\%    & 96.1\%             & \textbf{95.3\%}    & \radd{\textbf{91.1\%}}    & \radd{93.0\%}    & \textbf{938}      & 22020     \\ \hline \hline
square                   & 5.1E+07                                  & 93.0\%             & N/A       & 89.8\%              & N/A       & \radd{95.7\%}             & \radd{N/A}       & \textbf{13903}    & N/A       \\ \hline
\multirow{2}{*}{log2}    & 9.0E+00                                  & 94.1\%             & N/A       & 104.8\%             & N/A       & \radd{91.9\%}             & \radd{N/A}       & \textbf{37143}    & N/A       \\
                         & 8.4E+01                                  & 93.9\%             & N/A       & 103.7\%             & N/A       & \radd{90.9\%}             & \radd{N/A}       & \textbf{36754}    & N/A       \\ \hline
\multirow{2}{*}{sin}     & 5.0E+00                                  & 94.9\%             & N/A       & 110.2\%             & N/A       & \radd{94.6\%}             & \radd{N/A}       & \textbf{1195}     & N/A       \\
                         & 3.2E+01                                  & 79.7\%             & N/A       & 96.7\%              & N/A       & \radd{72.0\%}             & \radd{N/A}       & \textbf{22293}    & N/A       \\ \hline
\multirow{2}{*}{sqrt}    & 8.4E+01                                  & 81.2\%             & N/A       & 81.3\%              & N/A       & \radd{77.7\%}             & \radd{N/A}       & \textbf{400}      & N/A       \\
                         & 7.1E+03                                  & 62.6\%             & N/A       & 62.9\%              & N/A       & \radd{56.4\%}             & \radd{N/A}       & \textbf{363}      & N/A       \\ \hline
\multicolumn{2}{c|}{Average of all}                                  & 89.3\%             & N/A       & 94.6\%             & N/A       & \radd{87.4\%}             & \radd{N/A}       & \textbf{7531}     & N/A       \\ \toprule
\end{tabular}
\vspace{-2.5em}
\end{table*}

\subsubsection{Experiments on EPFL Benchmarks}\label{subsubsec:res-epfl}
To show the scalability of our ALS flow,
we test it on the large EPFL arithmetic benchmarks.
The MUSCAT method cannot handle them,
so we do not compare our method with MUSCAT and we only compare our method with MECALS.
Similarly, the MaxED bounds are set to \radd{$\lfloor 2^{O/10}\rfloor$ and $\lfloor 2^{O/5} \rfloor$} for each benchmark,
where $O$ is the PO number of the benchmark.
To accelerate our flow,
we first apply the constant LACs to quickly simplify the circuit,
followed by the SASIMI LACs for further simplification.
Specifically, the constant LACs are first applied to the circuit POs (like truncation) until the MaxED bound is reached and then to the internal nodes of the circuit until the MaxED bound is reached.
Finally, the SASIMI LACs are applied until the MaxED bound is reached.
\radd{We emphasize that despite the above modification, the ALS flow is still based on the general framework 
presented in Fig.~\ref{fig:ALS_flow}.
This modification is introduced only to reduce runtime on large benchmarks,
where directly considering all constant and SASIMI LACs in each iteration is computationally expensive ($>$ 24 hours).
With the modification,
we reduce the number of LACs to be checked while still following the same ALS flow in Fig.~\ref{fig:ALS_flow}.}

Table~\ref{tab:res_maxed_epfl} compares the area ratio, delay ratio,\radd{ power ratio,} and runtime of our flow and MECALS on the EPFL arithmetic benchmarks under the MaxED constraint.
We can see that our method can handle all benchmarks in the table with an average runtime of $7531$ seconds,
while MECALS cannot handle the benchmarks \tit{log2}, \tit{sin}, \tit{sqrt}, and \tit{square} (under $5.1 \times 10^7$ MaxED bound) in 24 hours.
\radd{For the benchmarks that both our method and MECALS can handle (the top part of the table),
our method further reduces the area by $2.9\%$ and power by $1.9\%$ on average with a small delay overhead,
while being $23.5\times$ faster than MECALS on average.
Notably, for the benchmark \tit{add128},
under the two MaxED bounds of $7.6\times 10^3$ and $5.8\times 10^7$,
our method speeds up by $65.8\times$ and $296.7\times$, respectively, compared to MECALS and
reduces area by $7.3\%$ and $6.5\%$, respectively,
and power by $5.4\%$ and $4.1\%$, respectively.}
Only for the benchmark \tit{bar} under the MaxED bound of $5.1\times 10^7$,
our method is slightly worse than MECALS in terms of the area ratio.
However, our method is far more efficient than MECALS with a competitive delay ratio.
\radd{Moreover, across the cases where both methods succeed (the top part of Table~\ref{tab:res_maxed_epfl}),
our method finishes after an average of 3.8 iterations,
and the average number of LACs applied per iteration is 11.4.
In contrast, MECALS needs an average of 37.6 iterations
with only one LAC applied per iteration.
This again shows the efficiency of our method.}

\vspace{-1em}
\subsection{Experiments Under the MaxHD Constraint}\label{subsec:res-maxhd}

\begin{table*}[!htbp]
\caption{Comparison of our ALS flow with the MECALS method on the ISCAS\radd{ and EPFL arithmetic} benchmarks under the MaxHD constraint.
\radd{\tbf{Bold} entries indicate smaller area, delay, and power ratios, or shorter runtime. 
N/A means that MECALS cannot obtain the final approximate circuit in 24 hours.}
OOM means that MECALS terminates due to out of memory.}\label{tab:res_maxhd_iscas}
\vspace{-1em}
\centering

\begin{tabular}{cr|rr|rr|rr|rr}
\bottomrule
\multirow{2}{*}{Circuit} & \multirow{2}{*}{\makecell{MaxHD\\bound}} & \multicolumn{2}{c|}{Area ratio}    & \multicolumn{2}{c|}{Delay ratio}     & \multicolumn{2}{c|}{Power ratio} & \multicolumn{2}{c}{Runtime/s} \\
                         &             & Ours            & MECALS          & Ours             & MECALS           & \radd{Ours}               & \radd{MECALS}     & Ours          & MECALS        \\ \hline
\multirow{2}{*}{c880}    & 2           & \textbf{88.3\%} & 97.3\%          & \textbf{100.5\%} & 105.6\%          & \radd{\textbf{85.9\%}}    & \radd{97.6\%}     & \textbf{0.9}  & 1.9           \\
                         & 5           & \textbf{67.5\%} & 92.4\%          & 113.1\%          & \textbf{106.8\%} & \radd{\textbf{55.6\%}}    & \radd{95.3\%}     & \textbf{1.1}  & 2.8           \\ \hline
\multirow{2}{*}{c1355}   & 3           & \textbf{10.1\%} & 10.4\%          & 3.4\%            & 3.4\%            & \radd{\textbf{8.1\%}}     & \radd{8.3\%}     & \textbf{0.6}  & 73            \\
                         & 6           & \textbf{9.1\%}  & 10.8\%          & 3.4\%            & 3.4\%            & \radd{\textbf{7.3\%}}     & \radd{8.6\%}     & \textbf{0.4}  & 17            \\ \hline
\multirow{2}{*}{c1908}   & 2           & \textbf{89.9\%} & 97.3\%          & \textbf{89.5\%}  & 101.1\%          & \radd{\textbf{89.7\%}}    & \radd{101.0\%}    & \textbf{1.0}  & 5.0           \\
                         & 5           & 73.6\%          & \textbf{65.5\%} & \textbf{84.3\%}  & 86.6\%           & \radd{\textbf{62.2\%}}    & \radd{75.3\%}    & \textbf{1.5}  & 53            \\ \hline
\multirow{2}{*}{c2670}   & 14          & \textbf{64.2\%} & 84.9\%          & \textbf{62.9\%}  & 99.6\%           & \radd{\textbf{66.2\%}}    & \radd{83.4\%}    & 29            & \textbf{26}   \\
                         & 28          & 42.3\%          & \textbf{41.2\%} & \textbf{36.9\%}  & 54.4\%           & \radd{\textbf{42.8\%}}    & \radd{44.9\%}    & \textbf{32}   & 80            \\ \hline
\multirow{2}{*}{c3540}   & 2           & \textbf{93.8\%} & 95.5\%          & \textbf{92.8\%}  & 109.1\%          & \radd{\textbf{94.2\%}}    & \radd{97.6\%}    & \textbf{1.3}  & 64            \\
                         & 4           & \textbf{92.1\%} & 95.6\%          & \textbf{98.7\%}  & 102.8\%          & \radd{\textbf{91.8\%}}    & \radd{94.6\%}    & \textbf{1.6}  & 105           \\ \hline
\multirow{2}{*}{c5315}   & 12          & \textbf{92.7\%} & 96.5\%          & 104.2\%          & \textbf{103.5\%} & \radd{\textbf{94.0\%}}    & \radd{96.8\%}    & \textbf{25}   & 158           \\
                         & 24          & \textbf{84.6\%} & 90.5\%          & \textbf{93.8\%}  & 99.6\%           & \radd{\textbf{85.7\%}}    & \radd{91.7\%}    & \textbf{34}   & 877           \\ \hline
\multirow{2}{*}{c7552}   & 10          & \textbf{41.2\%} & 95.8\%          & \textbf{77.6\%}  & 99.8\%           & \radd{\textbf{40.6\%}}    & \radd{96.2\%}    & \textbf{16}   & 193           \\
                         & 21          & \textbf{32.7\%} & 84.2\%          & \textbf{68.2\%}  & 100.1\%          & \radd{\textbf{29.6\%}}    & \radd{83.9\%}    & \textbf{29}   & 627           \\ \hline
                                & \radd{12} & \radd{\textbf{92.0\%}} & \radd{92.4\%} & \radd{113.2\%}          & \radd{\textbf{96.3\%}}  & \radd{\textbf{89.3\%}} & \radd{90.2\%}  & \radd{\textbf{8.2}}  & \radd{9.7}           \\
\multirow{-2}{*}{\radd{add128}} & \radd{25} & \radd{\textbf{83.7\%}} & \radd{88.9\%} & \radd{101.7\%}          & \radd{\textbf{101.3\%}} & \radd{\textbf{80.4\%}} & \radd{84.8\%}  & \radd{35}            & \radd{\textbf{16}}   \\ \hline
                                & \radd{12} & \radd{\textbf{97.0\%}} & \radd{99.5\%} & \radd{101.0\%}          & \radd{\textbf{98.8\%}}  & \radd{\textbf{94.5\%}} & \radd{99.2\%}  & \radd{\textbf{27}}   & \radd{13761}         \\
\multirow{-2}{*}{\radd{bar}}    & \radd{25} & \radd{\textbf{90.8\%}} & \radd{98.5\%} & \radd{99.8\%}           & \radd{\textbf{95.8\%}}  & \radd{\textbf{85.0\%}} & \radd{95.0\%}  & \radd{\textbf{35}}   & \radd{18128}         \\ \hline

                                & \radd{3} & \radd{100.0\%} & OOM & \radd{100.0\%}          & OOM  & \radd{100.0\%} & OOM  & \radd{\textbf{22045}}   & OOM         \\
\multirow{-2}{*}{log2}    & \radd{6} & \radd{100.0\%} & OOM & \radd{100.0\%}          & OOM  & \radd{100.0\%} & OOM  & \radd{\textbf{24133}}   & OOM         \\ \hline

                                & \radd{13} & \radd{\textbf{86.6\%}} & \radd{94.2\%} & \radd{\textbf{74.7\%}}  & \radd{83.1\%}           & \radd{\textbf{82.7\%}} & \radd{89.9\%}  & \radd{260}           & \radd{\textbf{38}}   \\                        
\multirow{-2}{*}{\radd{max}}    & \radd{26} & \radd{\textbf{77.0\%}} & \radd{92.7\%} & \radd{\textbf{64.7\%}}  & \radd{80.3\%}           & \radd{\textbf{71.7\%}} & \radd{86.7\%}  & \radd{172}           & \radd{\textbf{79}}   \\ \hline
                                & \radd{12} & \radd{\textbf{95.9\%}} & \radd{99.1\%} & \radd{98.9\%}           & \radd{98.9\%}           & \radd{\textbf{97.1\%}} & \radd{100.9\%} & \radd{\textbf{357}}  & \radd{13932}         \\
\multirow{-2}{*}{\radd{mult64}} & \radd{25} & \radd{\textbf{94.9\%}} & \radd{98.6\%} & \radd{\textbf{95.2\%}}  & \radd{97.9\%}           & \radd{\textbf{96.4\%}} & \radd{100.2\%} & \radd{\textbf{2192}} & \radd{22736}         \\ \hline
                                & \radd{2}  & \radd{\textbf{93.6\%}} & \radd{97.9\%} & \radd{104.9\%}          & \radd{\textbf{101.5\%}} & \radd{\textbf{93.7\%}} & \radd{97.9\%}  & \radd{3002}          & \radd{\textbf{1436}} \\
\multirow{-2}{*}{\radd{sin}}    & \radd{5}  & \radd{\textbf{92.8\%}} & \radd{97.1\%} & \radd{\textbf{100.9\%}} & \radd{102.0\%}          & \radd{\textbf{92.7\%}} & \radd{96.9\%}  & \radd{4832}          & \radd{\textbf{2062}} \\ \hline
                                & \radd{6}  & \radd{81.2\%}          & \radd{N/A}    & \radd{82.2\%}           & \radd{N/A}              & \radd{78.1\%}          & \radd{N/A}     & \radd{\textbf{560}}  & \radd{N/A}           \\
\multirow{-2}{*}{\radd{sqrt}}   & \radd{12} & \radd{65.2\%}          & \radd{N/A}    & \radd{64.8\%}           & \radd{N/A}              & \radd{58.9\%}          & \radd{N/A}     & \radd{\textbf{753}}  & \radd{N/A}           \\ \hline
                                & \radd{12} & \radd{\textbf{92.9\%}} & \radd{99.0\%} & \radd{\textbf{95.1\%}}  & \radd{95.5\%}           & \radd{\textbf{94.7\%}} & \radd{98.4\%}  & \radd{\textbf{284}}  & \radd{37073}         \\
\multirow{-2}{*}{\radd{square}} & \radd{25} & \radd{\textbf{91.8\%}} & \radd{98.5\%} & \radd{93.4\%}           & \radd{\textbf{92.4\%}}  & \radd{\textbf{92.6\%}} & \radd{97.7\%}  & \radd{\textbf{2075}} & \radd{45644}         \\ \hline
\multicolumn{2}{c|}{\radd{Average w/o log2 \& sqrt}}& \radd{\textbf{75.8\%}} & \radd{85.2\%} & \radd{\textbf{83.6\%}}  & \radd{89.2\%}           & \radd{\textbf{74.0\%}} & \radd{85.1\%}  & \radd{\textbf{517}}  & \radd{6046} \\ \toprule
\end{tabular}
\vspace{-2em}
\end{table*}

This set of experiments approximates the ISCAS85\radd{ and EPFL arithmetic} benchmarks in Table~\ref{tab:benchmark} under the MaxHD constraint.
We compare our ALS flow with MECALS.
We modify the open-source codes of MECALS to support the MaxHD constraint
and run both methods on the same platform for fair comparison.
MUSCAT is not compared in this experiment as its open-source code does not support the MaxHD constraint.
The MaxHD bounds are set to \radd{$\lfloor O/10 \rfloor$ and $\lfloor O/5 \rfloor$} for each benchmark,
where $O$ is the number of POs of the benchmark.
\radd{\fvdel{Under these MaxHD bounds, the \tit{log2} benchmark cannot be approximated by either our method or MECALS, so it is excluded from this experiment.}%
For the \tit{max} benchmark, we set the parameter $K$ in the sorting strategy in Section~\ref{subsec:sorter} to $1000$ to achieve better approximate circuits with lower hardware costs,
while for the others, $K$ still keeps the default value of $100$.}
Similar as in Section~\ref{subsubsec:res-epfl}, 
to accelerate our flow,
the constant LACs are first applied until the MaxHD bound is reached,
and then the SASIMI LACs are applied until the MaxHD bound is reached.

\radd{Table~\ref{tab:res_maxhd_iscas} compares the area ratio, delay ratio, power ratio, 
and runtime of our flow and MECALS.
Our method completes all benchmarks within 7 hours,
while MECALS fails to process \tit{sqrt} within 24 hours.
For \tit{log2}, MECALS cannot obtain the final approximate circuit due to out of memory.
Under the given MaxHD bounds, our flow cannot apply any LAC to simplify \tit{log2} and therefore terminates after one iteration, with the one-round runtime reported in Table~\ref{tab:res_maxhd_iscas}.
Our method achieves smaller area and delay ratios on most benchmarks,
and smaller power ratios on all benchmarks.
Excluding \tit{log2} and \tit{sqrt}, our method achieves average area, delay, and power ratios of $75.8\%$, $83.6\%$, and $74.0\%$, respectively, with an average runtime of 517 seconds.
Compared to MECALS, our method reduces area, delay, and power by $9.4\%$, $5.6\%$, and $11.1\%$ on average, respectively.
Notably, for benchmark \tit{c7552}, both area and power savings exceed $50\%$.}
For benchmark \tit{c1908} under the MaxHD bound of $5$ and benchmark \tit{c2670} under the MaxHD bound of $28$,
although our method is worse than MECALS in terms of the area ratio,
it reduces delay and power consumption.
\radd{Moreover, our method is more efficient than MECALS,
accelerating over MECALS by 11.7$\times$ on average.
The efficiency is attributed to the reduced number of iterations by our method.
Across all benchmarks in Table~\ref{tab:res_maxhd_iscas},
our method finishes after an average of 5.6 iterations,
and the average number of LACs applied per iteration is 21.3.
In contrast, MECALS needs an average of 48.9 iterations
with only one LAC applied per iteration.}
\subsection{Effectiveness of Simulation\radd{ and Parameter Study}}\label{subsec:res-ablation}

\radd{This section evaluates the effectiveness of simulation in our ALS flow.
It also studies the impact of several important parameters in our method on the synthesis quality and runtime.}

\begin{table*}[!htbp]
\caption{Comparison of our ALS flows with and without the simulation-guided LAC pruning under the MaxED constraint.
Only constant LACs are used in this experiment.
\radd{\tbf{Bold} entries indicate smaller area or delay ratios, shorter runtime, or fewer SAT problems solved.}}\label{tab:res_pruner}
\vspace{-1em}
\centering
\tabcolsep=2.0pt
\begin{tabular}{@{}cr|rr|rr|rr|rr@{}}
\bottomrule
\multirow{2}{*}{Circuit} & \multirow{2}{*}{\makecell{MaxED\\bound}} & \multicolumn{2}{c|}{Area ratio}  & \multicolumn{2}{c|}{Delay ratio}    & \multicolumn{2}{c|}{Runtime/s} & \multicolumn{2}{c}{Solved \#SAT} \\
& & With pruning & W/o pruning & With pruning & W/o pruning & With pruning & W/o pruning  & With pruning  & W/o pruning \\ \hline

\multirow{2}{*}{absdiff} & 1    & 81.7\%          & 81.7\%          & 88.7\%          & 88.7\%           & \textbf{0.1}  & 1.4  & \textbf{10}  & 518  \\
                         & 3    & 78.0\%          & \textbf{53.6\%} & \textbf{77.9\%} & 94.2\%           & \textbf{0.1}  & 1.7  & \textbf{10}  & 492  \\ \hline
\multirow{2}{*}{add8}    & 1    & 86.1\%          & 86.1\%          & 91.2\%          & 91.2\%           & \textbf{0.03} & 0.1  & \textbf{0}   & 122  \\
                         & 3    & 72.2\%          & 72.2\%          & 77.0\%          & 77.0\%           & \textbf{0.05} & 0.2  & \textbf{4}   & 212  \\ \hline
\multirow{2}{*}{add32}   & 9    & 72.0\%          & 72.0\%          & 98.3\%          & 98.3\%           & \textbf{0.2}  & 5.7  & \textbf{18}  & 952  \\
                         & 97   & 63.5\%          & 63.5\%          & 88.3\%          & 88.3\%           & \textbf{0.4}  & 4.9  & \textbf{39}  & 874  \\ \hline
\multirow{2}{*}{binsqrd} & 3    & 77.7\%          & 77.7\%          & 94.2\%          & 94.2\%           & \textbf{52}   & 1233 & \textbf{20}  & 6180 \\
                         & 12   & 77.0\%          & 77.5\%          & 94.2\%          & 94.2\%           & \textbf{343}  & 1362 & \textbf{67}  & 6134 \\ \hline
\multirow{2}{*}{buttfly} & 10   & 78.2\%          & 78.2\%          & 110.7\%         & 110.7\%          & \textbf{0.3}  & 1.1  & \textbf{36}  & 1046 \\
                         & 111  & 75.5\%          & 75.5\%          & 113.2\%         & 113.2\%          & \textbf{0.3}  & 1.1  & \textbf{58}  & 1028 \\ \hline
\multirow{2}{*}{mac}     & 1    & 92.6\%          & 92.6\%          & 100.8\%         & 100.8\%          & \textbf{0.1}  & 1.3  & \textbf{4}   & 554  \\
                         & 2    & 85.7\%          & 85.7\%          & 95.5\%          & 95.5\%           & \textbf{0.1}  & 1.2  & \textbf{9}   & 532  \\ \hline
\multirow{2}{*}{mult8}   & 3    & 73.9\%          & 73.9\%          & 85.6\%          & 85.6\%           & \textbf{0.4}  & 71   & \textbf{4}   & 2560 \\
                         & 9    & 72.5\%          & 72.3\%          & \textbf{80.5\%} & 81.0\%           & \textbf{0.8}  & 106  & \textbf{27}  & 2532 \\ \hline
\multirow{2}{*}{mult16}  & 9    & 99.3\%          & 99.3\%          & 100.5\%         & \textbf{100.0\%} & \textbf{3.5}  & 1208 & \textbf{15}  & 7886 \\
                         & 84   & \textbf{93.6\%} & 93.7\%          & \textbf{91.7\%} & 91.8\%           & \textbf{57}   & 1357 & \textbf{124} & 7700 \\ \hline
\multirow{2}{*}{mult32}  & 84   & 98.4\%          & 98.4\%          & \textbf{93.5\%}          & 95.2\%           & \textbf{162}  & 14549& \textbf{111} & 15872 \\
                         & 7131 & \textbf{94.5\%} & 94.7\%          & \textbf{94.8\%}          & 101.2\%          & \textbf{669}  & 14437& \textbf{417} & 15398 \\ \hline
\multicolumn{2}{c|}{Average}     & 81.8\%          & \textbf{80.5\%} & \textbf{93.2\%} & 94.5\%           & \textbf{72}   & 1909         & \textbf{54}  &   3922   \\ \toprule
\vspace{-3em}
\end{tabular}
\end{table*}

\subsubsection{Effectiveness of Simulation-Guided LAC Pruning}\label{subsec:res-sim-effect}

To show the effectiveness of the simulation-guided LAC pruning (see Section~\ref{subsec:prune}) in our ALS flow,
we conduct an ablation study on the arithmetic circuits used in MECALS in Table~\ref{tab:benchmark} under the MaxED constraint.
We compare our flows with and without the simulation-guided LAC pruning.
For the flow without the pruning,
we evaluate all candidate LACs in each iteration and select the first $K$ valid LACs to simplify the circuit.
Similar to the previous experiments,
we choose MaxED bounds of \rdel{$\lfloor 2^{0.1O}\rfloor$ and $\lfloor 2^{0.2O}\rfloor$}\radd{$\lfloor 2^{O/10}\rfloor$ and $\lfloor 2^{O/5}\rfloor$} for each benchmark,
where $O$ is the number of POs of the benchmark.
Since the flow without the pruning is very slow,
we only use the simple constant LACs in this experiment to ensure that the experiment finishes in a reasonable time.

Table~\ref{tab:res_pruner} compares our ALS flows with and without the simulation-guided LAC pruning under the MaxED constraint
in terms of the area ratio, delay ratio, runtime, and the number of SAT problems solved in the flow.
We can see that with pruning applied,
the number of SAT problems solved in our flow is dramatically reduced by 98.7\% on average,
leading to an average runtime reduction of 96.2\%.
Meanwhile, applying the pruning almost does not affect the area and delay ratios of the approximate circuits.
This is because the pruning just removes the invalid LACs according to the simulation results,
and the valid LACs are still preserved in the design space,
ensuring good qualities of the approximate circuits.
An exception is the benchmark \tit{absdiff} under the MaxED bound of $3$,
where the area ratio with the pruning is much larger than that without the pruning,
while the delay ratio with the pruning is much smaller than that without the pruning.
We believe that this is caused by an area-delay trade-off of the technology mapping process,
since the final approximate AIGs produced with and without the pruning before the technology mapping has similar size (\textit{i.e.}, $116$ with pruning versus $115$ without pruning) and the same depth (\textit{i.e.}, $12$).

\begin{figure*}[!htbp]
    \centering
    \includegraphics[width=1.0\textwidth]{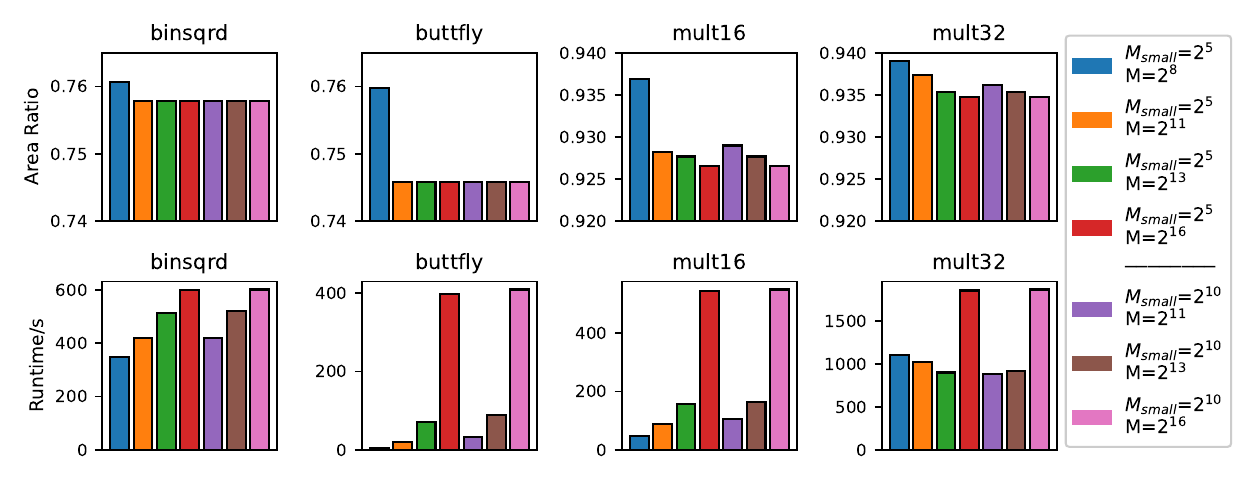}
    \vspace{-3em}
    \caption{\radd{Impact of simulation count on final area after applying our method and runtime of our method under $\lfloor 2^{O/5}\rfloor$ MaxED constraint. 
    Runtime data differs from that in Table~\ref{tab:res_maxed} since a different computer is used in this experiment.}}
    \label{fig:simlength_effect}
    \vspace{-1.5em}
\end{figure*}


\radd{%
\subsubsection{Impact of Simulation Count on Synthesis Quality and Runtime}\label{subsubsec:simlength_vs_quality}

We study the impact of the simulation count ($M_\tit{small}$ and $M$ in Section~\ref{subsubsec:prune_theory}) on both synthesis quality and runtime.
Recall that our method adopts a two-phase simulation-guided LAC pruning (see Section~\ref{subsubsec:prune_theory}).
First, $M_\tit{small}$ patterns are used for rough pruning of LACs,
followed by a larger set of $M$ patterns for further filtering.
We evaluate the impact of $M$ and $M_{\textit{small}}$ on four benchmarks, \tit{binsqrd, buttﬂy, mult16}, and \tit{mult32}, under a ﬁxed MaxED constraint of $\lfloor 2^{O/5}\rfloor$, 
where $O$ is the PO number of the benchmark.
For each benchmark,
we test seven $(M_\tit{small}, M)$ pairs:
$(2^5, 2^8)$, $(2^5, 2^{11})$, $(2^5, 2^{13})$, $(2^5, 2^{16})$, $(2^{10}, 2^{11})$, $(2^{10}, 2^{13})$, and $(2^{10}, 2^{16})$.
Note that $(2^{10}, 2^{13})$ is the default setting in all previous experiments.

Fig.~\ref{fig:simlength_effect} shows the impact of the simulation count on the final circuit area and runtime of our method.
For all benchmarks, under the same $M_{\textit{small}}$, 
the ﬁnal area decreases or remains unchanged as $M$ increases. This is because a larger $M$ can ﬁlter out more invalid LACs, 
thereby retaining more valid ones in the top $K$ candidates for further evaluation, 
which produces smaller approximate circuits.
Meanwhile,
under the same $M_{\textit{small}}$, 
the runtime generally increases with $M$, 
since more simulation patterns require more simulation time. 
An exception occurs for \tit{mult32} with $M_{\textit{small}}=2^5$, 
where runtime ﬁrst decreases and then increases. 
When both $M_{\textit{small}}$ and $M$ are small, pruning is ineffective, 
and many invalid LACs remain in the top $K$ LACs and must be verified by SAT solving, 
which dominates the runtime. 
As $M$ increases from $2^8$ to $2^{13}$, 
more invalid LACs are eliminated by simulation, 
leading to fewer SAT problems and reduced runtime. 
When $M$ further increases to $2^{16}$, 
the runtime rises again due to the long simulation time.
Overall, the default setting of $(M_\tit{small}, M)=(2^{10}, 2^{13})$, shown as brown bars in Fig.~\ref{fig:simlength_effect},
provides a good trade-off between the final area and runtime across the four benchmarks.}

\radd{%
\subsubsection{Impact of Simulation Count on SAT Solving Statistics}
We further study how the simulation count influences the SAT solving statistics.
In this study,
we fix $M_\tit{small}=2^5$ and vary $M\in\{2^8, 2^{11}, 2^{13}, 2^{16}\}$,
while keeping the other settings identical to those in the previous experiment in Section~\ref{subsubsec:simlength_vs_quality}.
Table~\ref{tab:simlength_vs_sat} shows,
for different $M$ values,
the total number of SAT problems solved, 
the percentages of SAT and UNSAT results, 
and the number of UNDEFINED results observed in the first two iterations of our method.
Since different $M$ values lead to different iteration numbers,
we restrict the comparison to the first two iterations for fairness.
As $M$ increases,
the percentage of SAT results generally decreases while that of UNSAT results generally increases.
This trend is expected because a larger $M$ prunes more invalid LACs,
which increases the likelihood that the preserved top $K$ LACs are valid.
As valid LACs correspond to UNSAT results when their maximum errors are formally checked,
the UNSAT ratio rises, 
while the SAT ratio decreases.
Moreover, UNDEFINED results are only observed in \tit{mult32} when $M=2^{11}, 2^{13},$ and $2^{16}$.
Due to the large size of \tit{mult32},
the SAT solver may fail to resolve some instances within the computation budget.}

\begin{table}[!htbp]
\centering
\caption{\radd{Impact of simulation count on SAT solving statistics in the first two iterations of our method under $\lfloor 2^{O/5}\rfloor$ MaxED constraint. 
$M_\tit{small}$ is fixed to $2^5$.}}
\label{tab:simlength_vs_sat}
\vspace{-1em}
\begin{tabular}{crrrrr}
\toprule
Circuit                  & $M$      & \makecell{All SAT\\calls} & \makecell{SAT\\percentage} & \makecell{UNSAT\\percentage} & \makecell{\#UNDEF\\count} \\ \midrule
\multirow{4}{*}{binsqrd} & $2^8$    & 98            & 42.9\%    & 57.1\%      & 0       \\
                         & $2^{11}$ & 60            & 3.3\%     & 96.7\%      & 0       \\
                         & $2^{13}$ & 60            & 3.3\%     & 96.7\%      & 0       \\
                         & $2^{16}$ & 60            & 3.3\%     & 96.7\%      & 0       \\ \midrule
\multirow{4}{*}{buttfly} & $2^8$    & 100           & 95.0\%    & 5.0\%       & 0       \\
                         & $2^{11}$ & 100           & 95.0\%    & 5.0\%       & 0       \\
                         & $2^{13}$ & 100           & 95.0\%    & 5.0\%       & 0       \\
                         & $2^{16}$ & 36            & 80.6\%    & 19.4\%      & 0       \\ \midrule
\multirow{4}{*}{mult16}  & $2^8$    & 17            & 17.6\%    & 82.4\%      & 0       \\
                         & $2^{11}$ & 15            & 0.0\%     & 100.0\%     & 0       \\
                         & $2^{13}$ & 15            & 0.0\%     & 100.0\%     & 0       \\
                         & $2^{16}$ & 15            & 0.0\%     & 100.0\%     & 0       \\ \midrule
\multirow{4}{*}{mult32}  & $2^8$    & 106           & 42.5\%    & 57.5\%      & 0       \\
                         & $2^{11}$ & 69            & 5.8\%     & 88.4\%      & 4       \\
                         & $2^{13}$ & 33            & 6.1\%     & 90.9\%      & 1       \\
                         & $2^{16}$ & 33            & 6.1\%     & 90.9\%      & 1       \\ \bottomrule
\end{tabular}
\vspace{-1em}
\end{table}

\begin{table}[!htbp]
\centering
\caption{\radd{Impact of number of top LACs selected ($K$) on final area, runtime, and number of iterations of our method under $\lfloor 2^{O/5}\rfloor$ MaxED constraint.
Runtime data differs from that in Table~\ref{tab:res_maxed} since a different computer is used in this experiment.}}
\label{tab:res-topk}
\vspace{-1em}
\begin{tabular}{crrrr}
\toprule
Circuit                  & $K$   & \makecell{Final\\Area/$\mu m^2$} & Runtime/s & \#Iterations \\ \midrule
\multirow{4}{*}{binsqrd} & 50    & 801.7    & 562       & 10    \\
                         & 100   & 797.5    & 445       & 7     \\
                         & 1000  & 795.6    & 851       & 5     \\
                         & 10000 & 790.3    & 927       & 5     \\ \midrule
\multirow{4}{*}{buttfly} & 50    & 127.2    & 48        & 7     \\
                         & 100   & 127.2    & 73        & 7     \\
                         & 1000  & 127.2    & 49        & 5     \\
                         & 10000 & 127.2    & 32        & 4     \\ \midrule
\multirow{4}{*}{mult16}  & 50    & 1327.3   & 149       & 12    \\
                         & 100   & 1316.2   & 142       & 7     \\
                         & 1000  & 1323.1   & 123       & 5     \\
                         & 10000 & 1328.9   & 729       & 4     \\ \midrule
\multirow{4}{*}{mult32}  & 50    & 5379.1   & 1240      & 19    \\
                         & 100   & 5353.0   & 866       & 10    \\
                         & 1000  & 5360.4   & 1377      & 7     \\
                         & 10000 & 5382.2   & 2924      & 4     \\ \bottomrule
\end{tabular}
\end{table}

\subsubsection{Impact of Number of Top LACs (Parameter $K$)}

Recall that during the LAC selection step shown in Fig.~\ref{fig:lac_select},
only the top $K$ LACs are kept for further evaluation based on their maximum error lower bounds and estimated area reductions.
We study the impact of $K$ on the final area, runtime, 
and iteration number of our method.
Specifically,
we test $K \in \{50, 100, 1000, 10000\}$ on the four benchmarks \tit{binsqrd, buttfly, mult16,} and \tit{mult32} under the MaxED constraint of $\lfloor 2^{O/5}\rfloor$.
The results are shown in Table~\ref{tab:res-topk}.

The impact of $K$ on the final area and runtime varies across benchmarks,
but a common related trend can be observed.
As $K$ increases,
the number of iterations decreases,
since more candidate LACs are evaluated and applied in each iteration.
However, a larger $K$ with fewer iterations does not necessarily yield shorter runtime,
because evaluating more LACs per iteration also requires solving more SAT problems per iteration.
Thus, the overall runtime is determined by the trade-off between the iteration count and the number of SAT calls per iteration.
For \tit{binsqrd}, \tit{mult16}, and \tit{mult32},
runtime decreases at first as $K$ increases,
but then rises again when $K$ further increases.
For \tit{buttfly},
runtime exhibits the opposite trend,
first increasing and then decreasing.

The effect of $K$ on the final circuit area also differs by benchmarks.
For \tit{binsqrd}, the final area consistently decreases with $K$.
For \tit{buttfly}, the final area remains the same for different $K$ values.
For \tit{mult16} and \tit{mult32}, the final area first decreases and then increases as $K$ grows.
One possible reason is that a small $K$ may exclude promising LACs from further evaluation,
leading to suboptimal results with larger area.
Conversely,
when $K$ is very large, 
too many LACs are applied in each iteration under the greedy selection strategy (see Fig.~\ref{fig:lac_select}).
This causes the circuit to deviate significantly from the current approximate circuit,
making the previously computed maximum error lower bounds of LACs inaccurate.
Guided by these inaccurate values,
some poor LACs with large errors may be applied, 
which eventually causes the flow to terminate prematurely
and produces a circuit with larger area.
\begin{figure*}[!htbp]
    \centering
    \subfigure[12-bit unsigned adder.]{
        \includegraphics[width=0.31\textwidth]{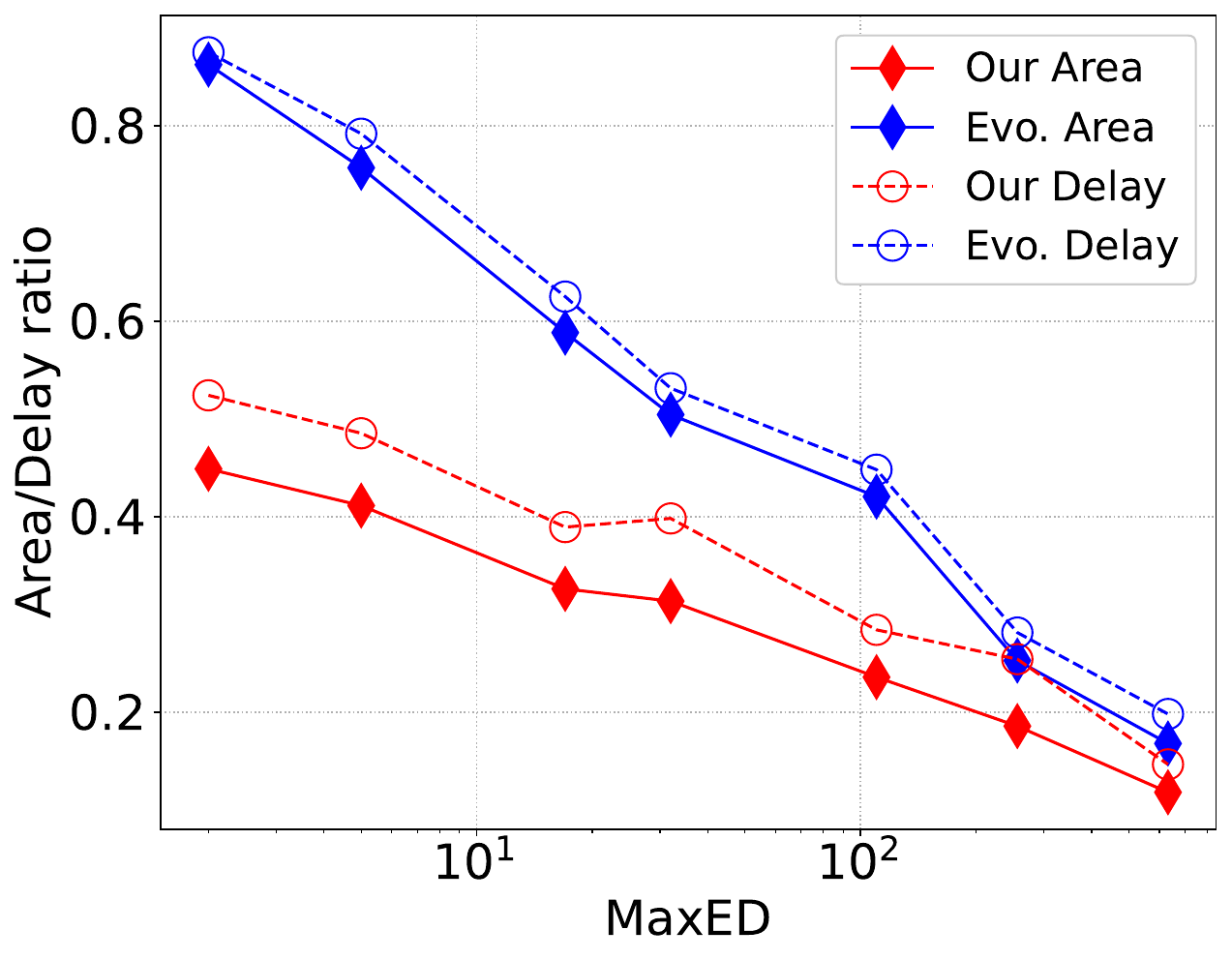} 
    }
    \subfigure[16-bit unsigned adder.]{
        \includegraphics[width=0.31\textwidth]{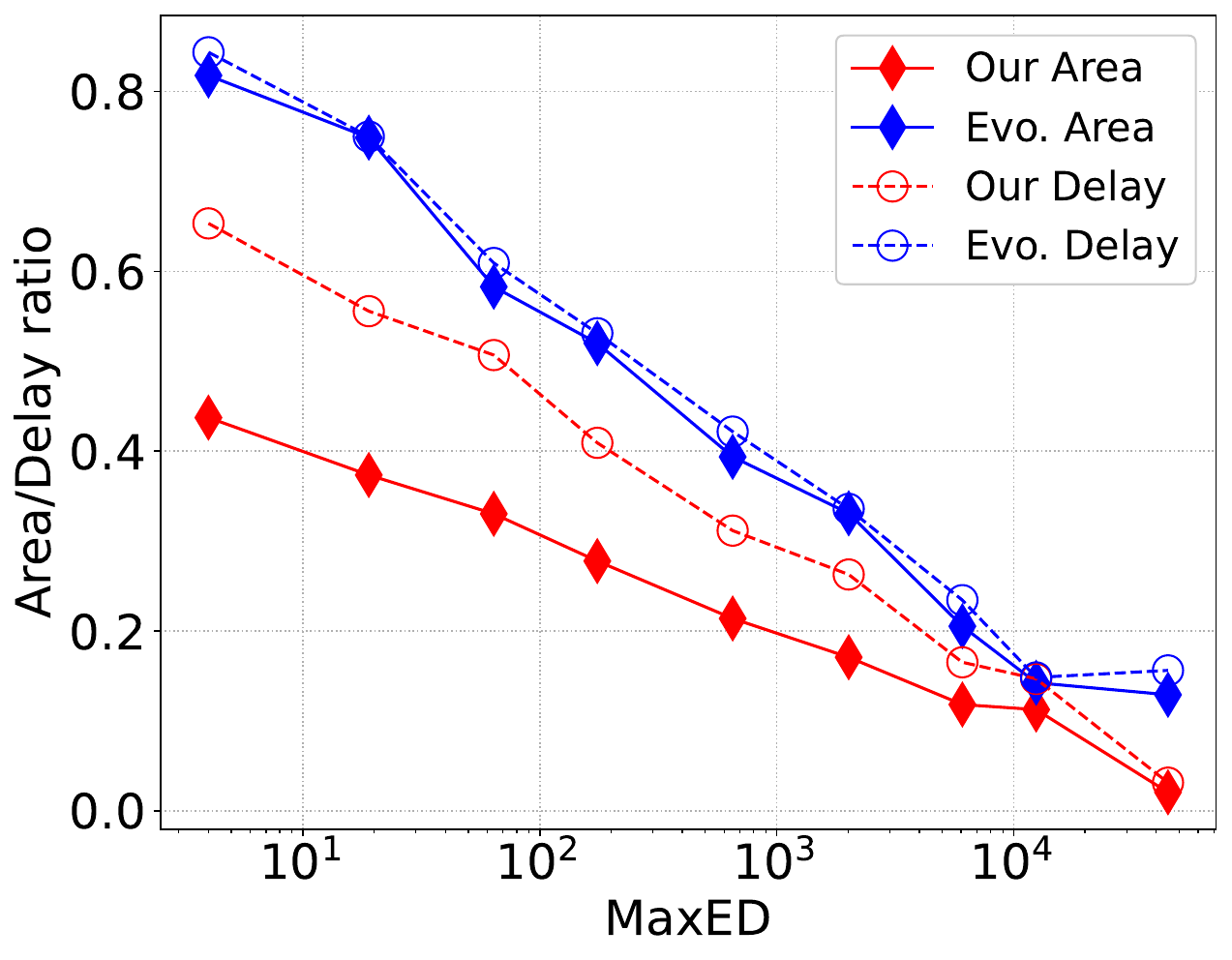} 
    }
    \subfigure[8-bit unsigned multiplier.]{
        \includegraphics[width=0.31\textwidth]{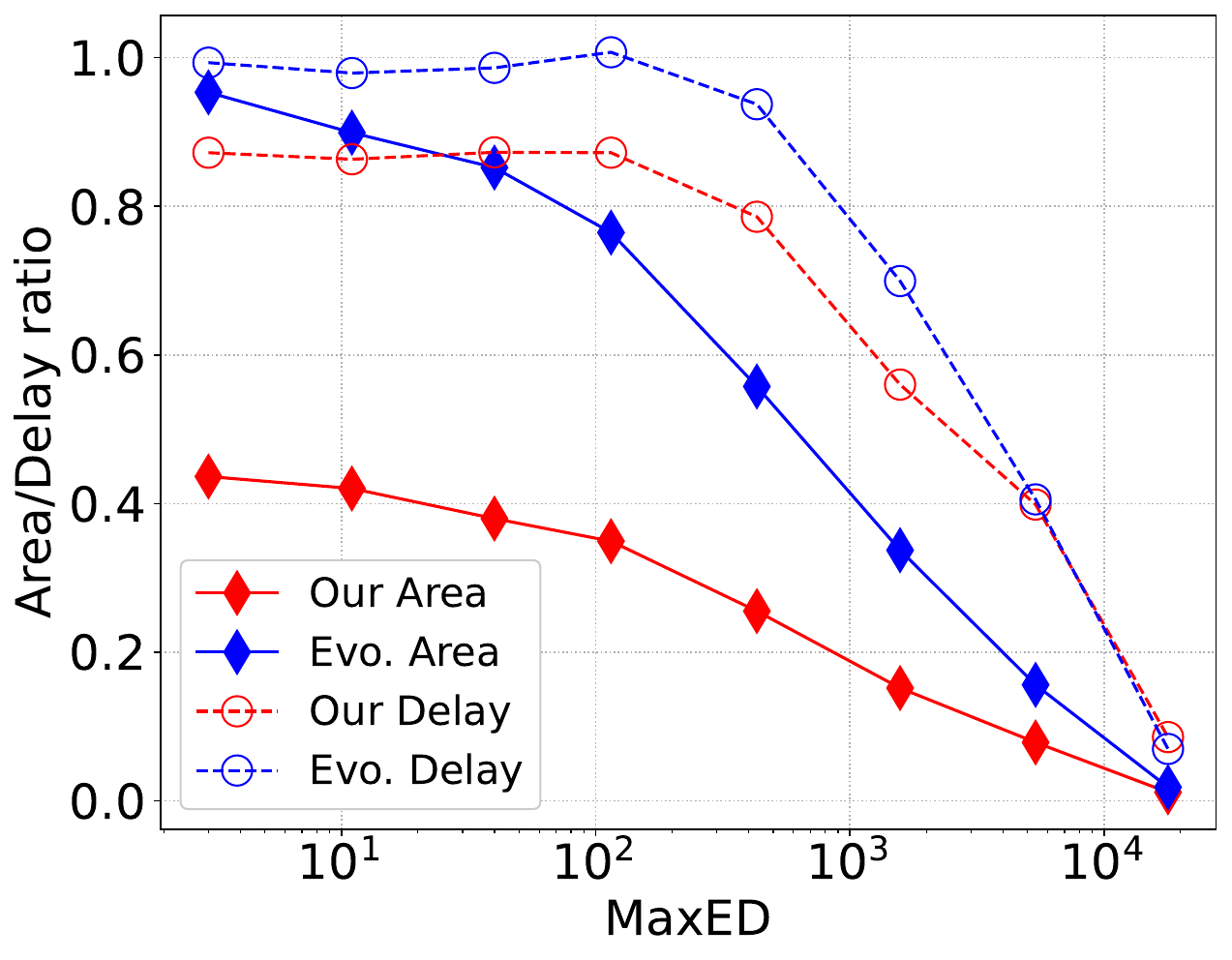} 
    }\\
    \vspace{-1em}
    \centering
    \subfigure[11-bit unsigned multiplier.]{
        \includegraphics[width=0.31\textwidth]{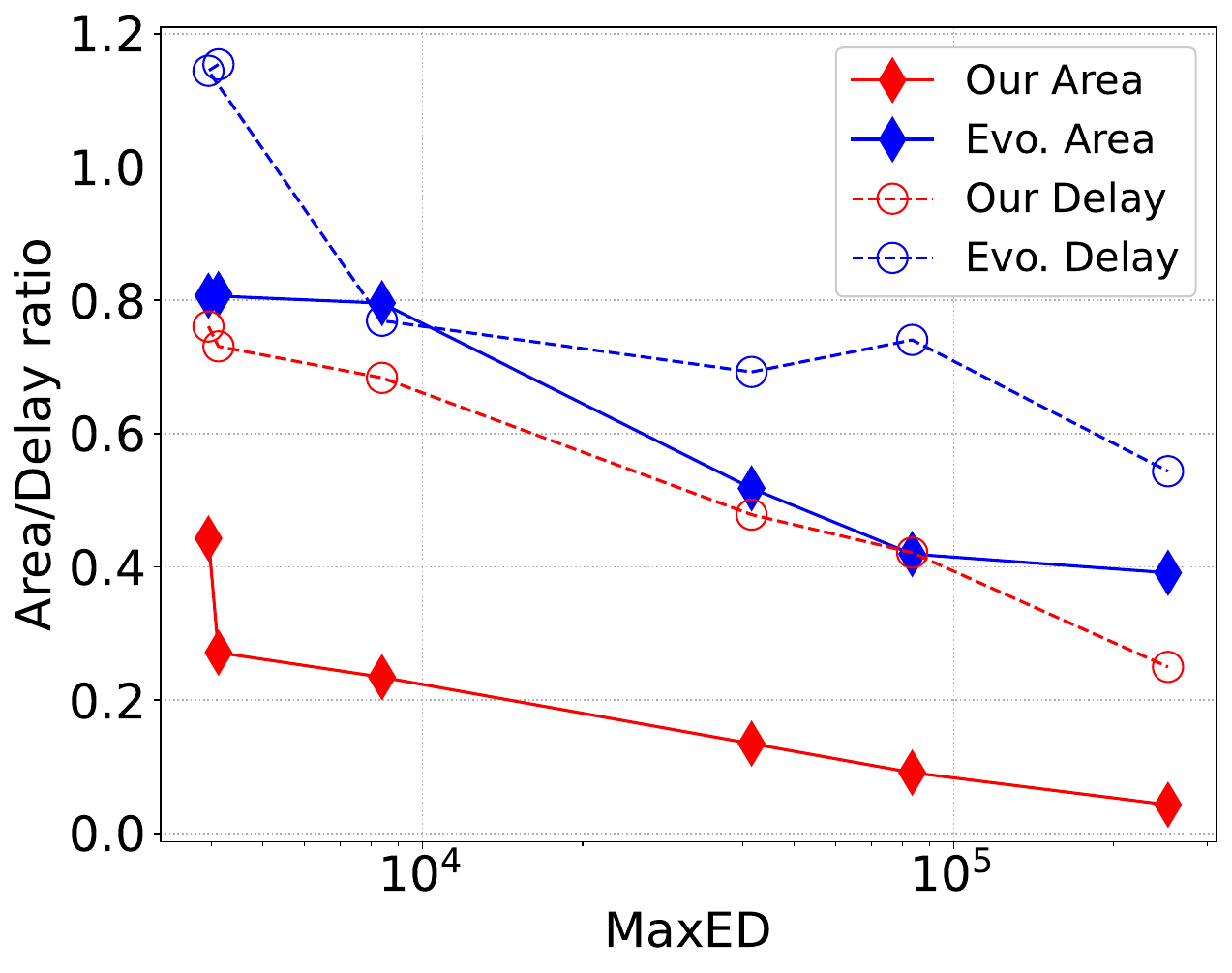} 
    }
    \subfigure[12-bit unsigned multiplier.]{
        \includegraphics[width=0.31\textwidth]{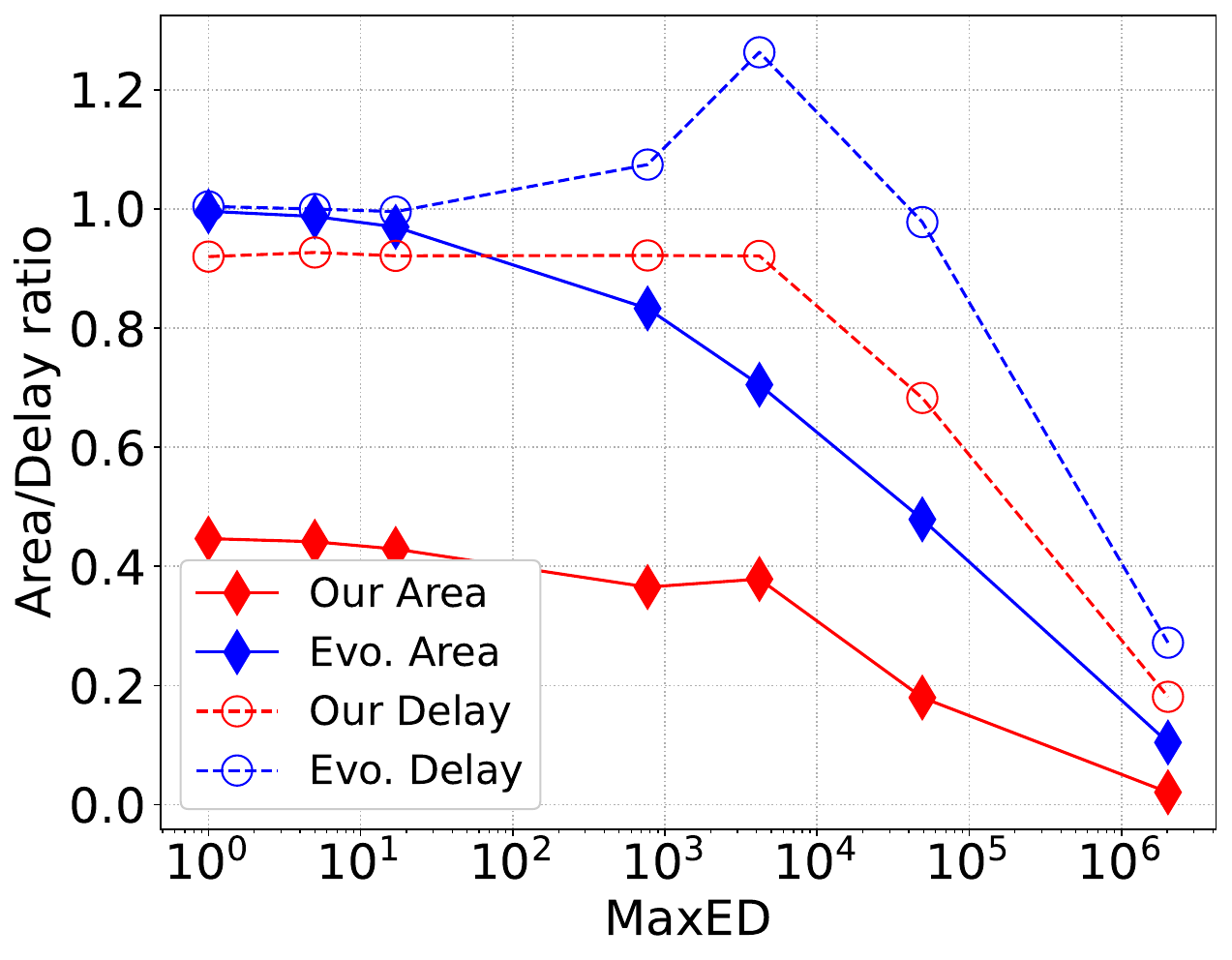} 
    }
    \subfigure[16-bit unsigned multiplier.]{
        \includegraphics[width=0.31\textwidth]{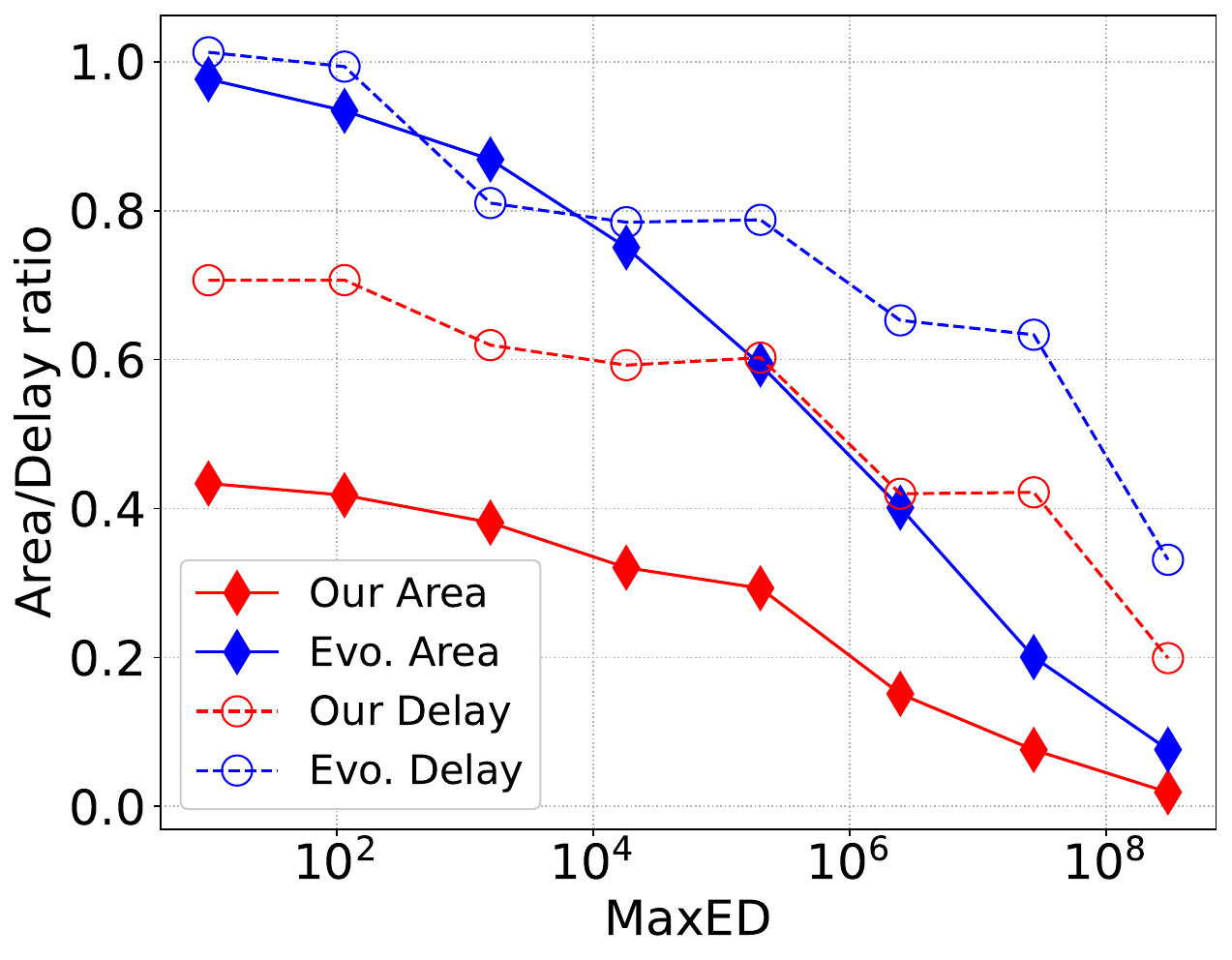} 
    }
    \vspace{-0.5em}
    \caption{Comparison between \cdel{the }approximate adders and multipliers synthesized by our flow and those from the EvoApproxLib under the MaxED constraint.}\label{fig:comp_evoapprox}
    \vspace{-1em}
\end{figure*}

\subsection{Comparison on Approximate Adder and Multiplier Designs}

Given the importance of approximate adders and multipliers,
especially in AI \cdel{hardware }accelerators,
we compare the approximate adders and multipliers synthesized by our ALS flow with those from the EvoApproxLib~\cite{mrazek2022evo} (version 2022),
a widely-used open-source library of approximate adders and multipliers.
The benchmarks compared are the 12-bit and 16-bit unsigned adders and the 8-bit, 11-bit, 12-bit, and 16-bit unsigned multipliers.
The approximate designs from the EvoApproxLib are those synthesized under the MaxED constraint.
Our ALS flow starts from the accurate circuits from the EvoApproxLib,
converts them to AIGs,
and applies the constant and SASIMI LACs to simplify the circuits under the MaxED constraint.
The MaxED bounds are set as the MaxEDs of the approximate circuits from the EvoApproxLib,
which means that the approximate circuits generated by our flow have the same or smaller MaxEDs compared to those in the EvoApproxLib.

Fig.~\ref{fig:comp_evoapprox} shows the comparison results,
where each sub-figure corresponds to an approximate adder or multiplier
and plots the area ratio-MaxED and the delay ratio-MaxED curves of the approximate circuits synthesized by our flow
and those from the EvoApproxLib.
Comparing the results from our flow (shown in red) with those from the EvoApproxLib (shown in blue),
we can see that under the same MaxED bound,
the approximate circuits synthesized by our ALS flow have much smaller area and delay ratios than those from the EvoApproxLib for all benchmarks.
Notably, when the MaxED is small,
there is a large gap between both the area and delay ratios of the approximate circuits synthesized by our flow and those from the EvoApproxLib.
This shows the effectiveness and practicality of our flow,
since reducing the hardware cost and delay under a small error bound is more challenging but important for real-world applications.
As the MaxED bound increases,
the area and delay ratios of the approximate circuits from both our flow and the EvoApproxLib decrease due to more approximation opportunities given by the large error bounds.
When the MaxED is large,
the improvement of our flow over the EvoApproxLib is reduced,
since the approximation opportunities are more abundant
and the ALS method used for producing the EvoApproxLib can also generate good approximate circuits in this case.

\section{Conclusion}\label{sec:concl}

This paper studies ALS under the maximum error constraint.
We propose to utilize logic simulation to guide the pruning of invalid LACs that violate the error constraint
and the selection of promising LACs for better circuit simplification.
By leveraging the simulation-guided techniques,
we further propose an efficient ALS flow that iteratively applies a set of promising LACs to approximate the input circuit.
The proposed flow can handle complex LACs and scale to large circuits with tens of thousands of gates.
The experimental results show that our ALS flow can achieve a better trade-off between error and hardware cost
compared to the state-of-the-art ALS methods.

\section*{Acknowledgement}

This work is supported 
\cdel{by the Swiss National Science Foundation Grant ``Supercool: Design methods and tools for superconducting electronics'' with funding number 200021\_1920981,}%
in part by the Synopsys Inc.,
by the National Natural Science Foundation of China under Grant 62574132, 
and by the Natural Science Foundation of Shanghai under Grant 25ZR1401189.
\radd{We thank the anonymous reviewers for their constructive feedback.}

\printbibliography

\vspace{-5.0em}
\begin{IEEEbiography}[{\includegraphics[width=0.90in,height=1.10in,clip,keepaspectratio]{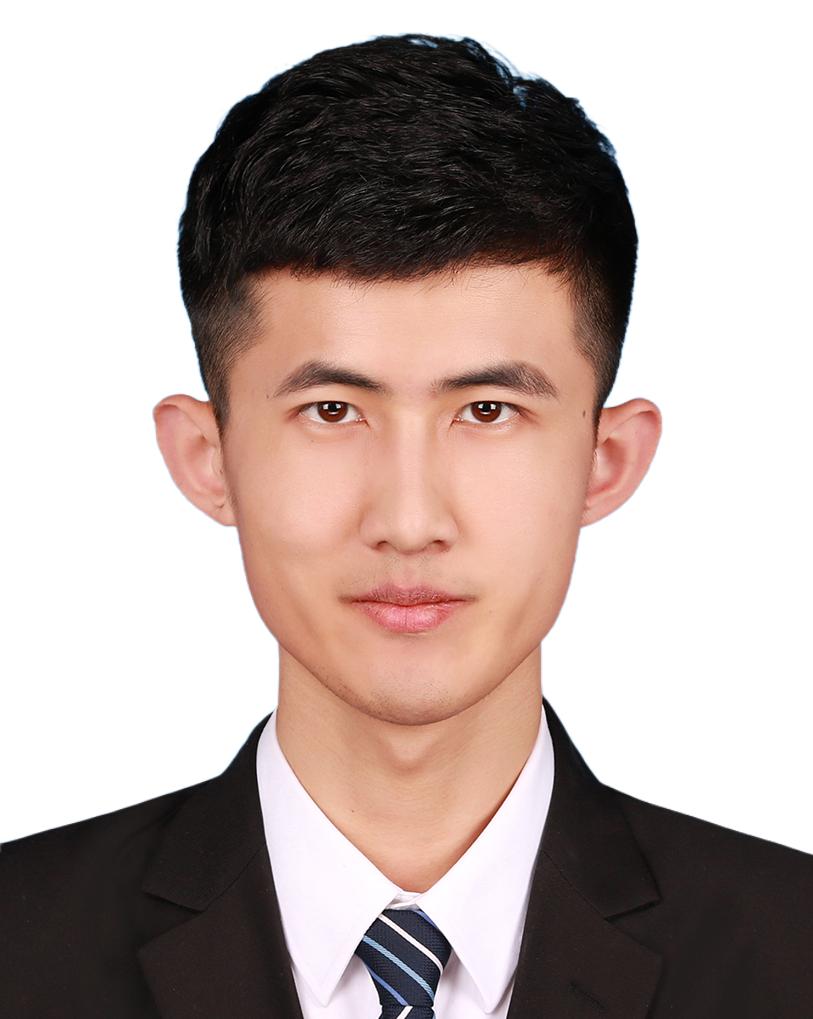}}]{Chang Meng}
    is a postdoctoral researcher at the Integrated Systems Laboratory, EPFL Lausanne, Switzerland.
    He received his Ph.D. degree in Electronic Science and Technology at Shanghai Jiao Tong University in 2023.
    His research interest is electronic design automation for emerging computing paradigms,
    especially the logic synthesis and verification of approximate computing circuits.
    His research work was nominated for the Best Paper Award at Design, Automation, and Test in Europe Conference (DATE).
\end{IEEEbiography}

\vspace{-5.0em}
\begin{IEEEbiography}[{\includegraphics[width=0.90in,height=1.10in,clip,keepaspectratio]{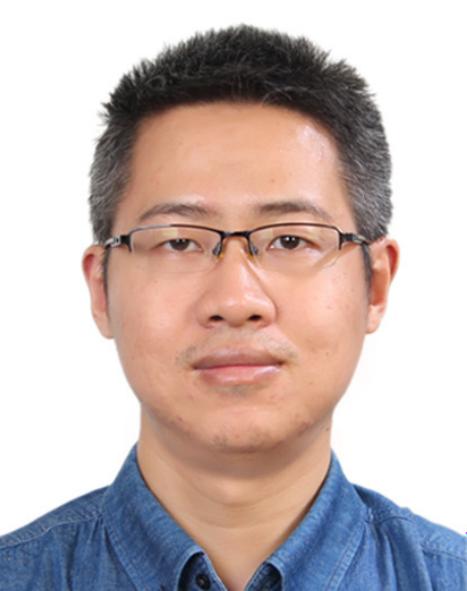}}]{Weikang Qian}
    is an associate professor in the Global College at Shanghai Jiao Tong University. 
    He received his Ph.D. degree in Electrical Engineering at the University of Minnesota in 2011 and his B.Eng. degree in Automation at Tsinghua University in 2006. His main research interests include electronic design automation and digital design for emerging computing paradigms. 
    His research works got the best student paper award at the International Workshop on Logic and Synthesis (IWLS) and the best paper nominations at the International Conference on Computer-Aided Design (ICCAD) and the Design, Automation, and Test in Europe Conference (DATE).
    He serves as an associate editor of the IEEE Transactions on Computer-Aided Design of Integrated Circuits and Systems. He is a senior member of IEEE.
\end{IEEEbiography}

\vspace{-5.0em}
\begin{IEEEbiography}[{\includegraphics[width=0.90in,height=1.10in,clip,keepaspectratio]{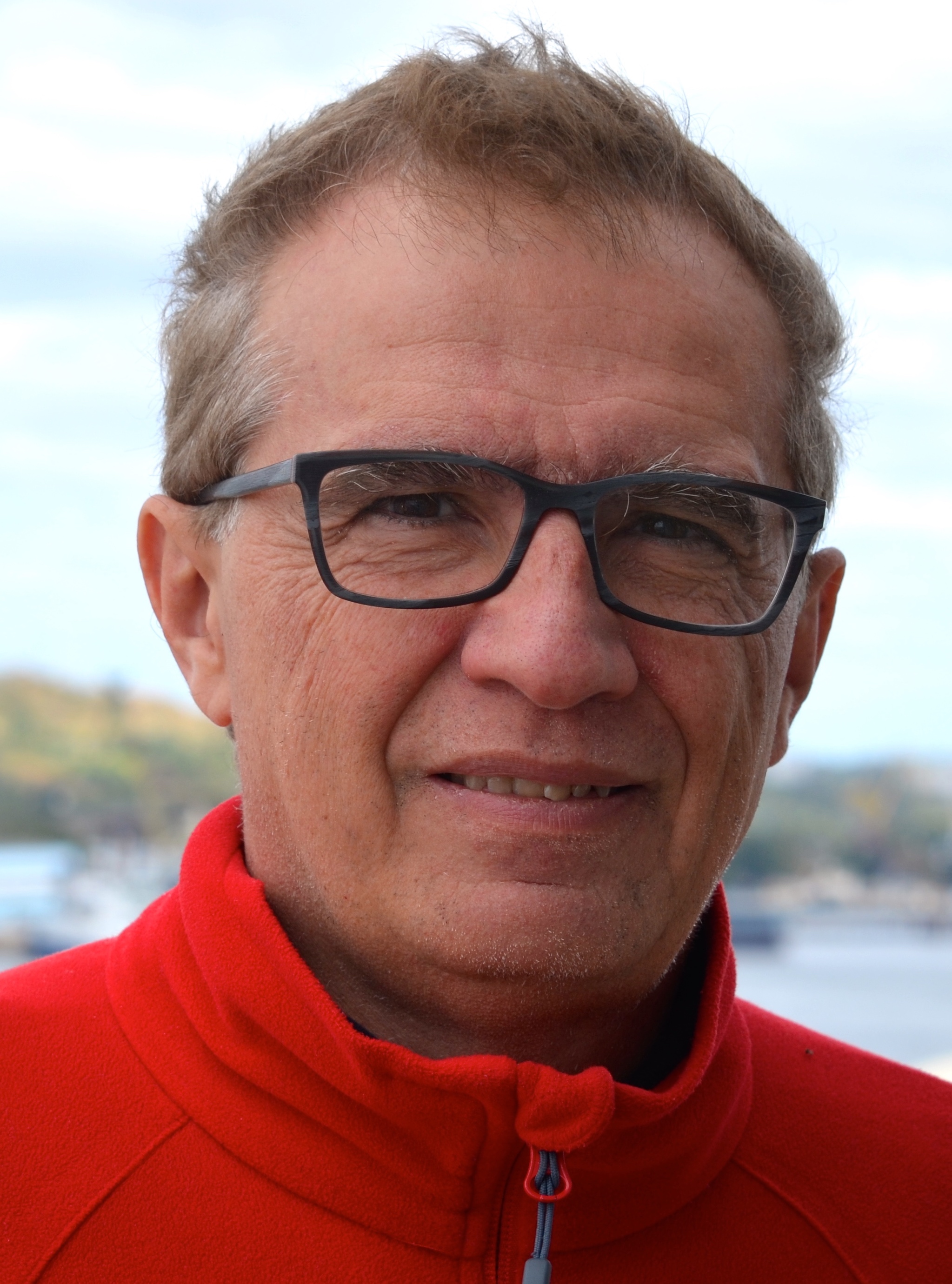}}]{Giovanni De Micheli}
    is Professor and Director of the Integrated Systems Laboratory and Scientific Director of the EcoCloud center at EPFL Lausanne, Switzerland.
    Previously, he was Professor of Electrical Engineering at Stanford University.
    He holds a Nuclear Engineer degree (Politecnico di Milano, 1979), 
    a M.S. and a Ph.D. degree in Electrical Engineering and Computer Science (University of California at Berkeley, 1980 and 1983).

    He is a Fellow of ACM, AAAS and IEEE, 
    a member of the Academia Europaea, of the Swiss Academy of Engineering Sciences, and International Honorary member of the American Academy of Arts and Sciences.
    His current research interests include several aspects of design technologies for integrated circuits and systems, such as synthesis for emerging technologies. 
    He is also interested in heterogeneous platform design including electrical components and biosensors, as well as in data processing of biomedical information.
    He is member of the Scientific Advisory Board of IMEC (Leuven, B) and STMicroelectronics.

    Professor De Micheli is the recipient of the 2025 IEEE Gustav Kirchhoff Award,
    the 2022 ESDA-IEEE/CEDA Phil Kaufman Award, 
    the 2019 ACM/SIGDA Pioneering Achievement Award, and several other awards.
\end{IEEEbiography}

\end{document}